\documentclass{ifacconf}

\usepackage{graphicx}      
\usepackage{natbib}        
\usepackage{amsmath}
\usepackage{amsfonts}
\usepackage{algorithm,algorithmic}
\newcommand{\w}{\mathbf{w}}
\newcommand{\x}{\mathbf{x}}

\renewcommand{\r}{\mathbf{r}}
\newcommand{\y}{\mathbf{y}}
\newcommand{\z}{\mathbf{z}}

\renewcommand{\v}{\mathbf{v}}
\newcommand{\f}{\mathbf{f}}
\newcommand{\g}{\mathbf{g}}

\newcommand{\gammab}{\boldsymbol{\gamma}}
\newcommand{\pib}{\boldsymbol{\pi}}
\newcommand{\Pib}{\boldsymbol{\Pi}}
\newcommand{\Thetab}{\boldsymbol{\Theta}}
\newcommand{\T}{^{T}}
\newcommand{\J}{\mathbf{J}}
\newcommand{\A}{\mathbf{A}}
\renewcommand{\L}{\mathbf{L}}
\newcommand{\K}{\mathbf{K}}
\newcommand{\R}{\mathbf{R}}

\newcommand{\TT}{\mathbf{T}}

\newcommand{\V}{\mathbf{V}}
\newcommand{\I}{\mathbf{I}}
\newcommand{\F}{\mathbf{F}}
\newcommand{\Q}{\mathbf{Q}}

\newcommand{\M}{\mathbf{M}}
\newcommand{\calK}{\mathcal{K}}
\newcommand{\calC}{\mathcal{C}}

\newcommand{\real}{\mathbb{R}}

\renewcommand{\P}[1]{\left(#1\right)}

\newcommand{\pd}[2][]{\frac{\partial #1}{\partial #2}}
\newcommand{\od}[2][]{\frac{{\rm d} #1}{{\rm d} #2}}

\renewcommand{\d}{{\rm d}}
\newcommand{\abs}[1]{|#1|}

\renewcommand{\half}{\frac{1}{2}}

\newcommand{\delt}[2][]{\ifthenelse{\equal{#1}{}} { \delta \mathbf{#2}}{\delta \mathbf{#2}_{#1}}}

\newtheorem{theorem}{Theorem}
\newtheorem{remark}{Remark}

\newtheorem{definition}{Definition}

\newtheorem{proposition}{Proposition}

\newcommand{\U}{\mathbf{U}}
\newcommand{\ddt}{ \od{t} }

\usepackage{verbatim}
\usepackage{color}

\begin{document}
\begin{frontmatter}

\title{Sparse Control for Dynamic Movement Primitives} 


\author[First]{Patrick M. Wensing} 
\author[First]{Jean-Jacques Slotine} 

\address[First]{Massachusetts Institute of Technology, 
   Cambridge, MA 02140 USA (e-mail: pwensing,jjs@mit.edu).}

\begin{abstract}                

This paper describes the use of spatially-sparse inputs to influence global changes in the behavior of Dynamic Movement Primitives (DMPs). The dynamics of DMPs are analyzed through the framework of contraction theory as networked hierarchies of contracting or transversely contracting systems. Within this framework, sparsely-inhibited rhythmic DMPs (SI-RDMPs) are introduced to both inhibit or enable rhythmic primitives through spatially-sparse modification of the DMP dynamics. SI-RDMPs are demonstrated in experiments to manage start-stop transitions for walking experiments with the MIT Cheetah. New analytical results on the coupling of oscillators with diverse natural frequencies are also discussed. 
\end{abstract}


\end{frontmatter}

\section{Introduction}

	There is a growing body of evidence that motor primitives may form the basis for a rich set of sensorimotor skills in humans and animals \citep{Bizzi94, Bizzi95, Hogan04, Hogan12}. From walking to grasping, the composition of primitive attractors could provide robustness as behaviors are generalized and recycled from past experience. Primitives may, in a sense, represent a compression of experience, capturing accumulations of knowledge that may be drawn on to simplify online control. This use of motor primitive techniques in biological systems would be well supported by the underlying nature of evolutionary change. Indeed, evolution necessarily proceeds through the accumulation of stable intermediate states \citep{Simon62}, building upon existing functional frameworks through stably layered complexity. 
	
	The use of dynamic movement primitives (DMPs) \citep{Schaal12} has sought to embody these principles for the development of sensorimotor skills in robotics.   Dynamic movement primitives are systems of coupled ordinary differential equations that represent a target attractor landscape for robot motion. The attractor landscapes can be learned through demonstration \citep{Ijspeert02} or crafted through manual design. The  landscapes of DMPs may represent attractors for a wide range of rhythmic and discrete movements \citep{Schaal06,Pastor09}.
	 	
	
	Rhythmic DMPs are closely related to the mimicry of biological Central Pattern Generators (CPGs) \citep{Marder01} within robotics \citep{Ijspeert08}. A hallmark of CPGs in biological systems is that a low-dimensional set of inputs can be used to orchestrate coordinated patterns of high-dimensional oscillatory motor control signals. Stable oscillations of Andronov-Hopf oscillators \citep{Chung10b} have been employed for pattern generation in bioinspired control of locomotion in air \citep{Chung10c} and water \citep{Chung10}. Stable phase oscillators \citep{Ajallooeian13} have been supplemented with sensorimotor feedback to stabilize quadrupedal locomotion \citep{Ijspeert13,Semini13}. Across these results, low-dimensional inputs are capable to smoothly reshape high-dimensional target behaviors for dynamic machines. 


	Despite the popularity of DMP/CPG frameworks, analysis of couplings between coordinated primitive modules has largely been lacking in the literature. Contraction analysis \citep{Slotine98} provides modular stability tools which may help to guide the architecture of more flexible and robust DMP/CPG frameworks. A preliminary analysis of discrete DMPs through  contraction theory was provided in \citep{Slotine06c}, with new analysis in this paper using transverse contraction theory \citep{Manchester14,Manchester14b}. Contracting systems are characterized by an exponential forgetting of initial conditions, providing a notion of stability without committing in advance to a particular trajectory. Such a notion is desirable from a practical standpoint, as success in situations form grasping a cup to running down a cliff are hardly characterized by unique solutions.

	The composition of primitive contracting systems suggests a promising approach for robust online synthesis from off-line knowledge \citep{Slotine98,Slotine06c,Slotine01,Manchester15}. As we will see, contracting systems provide an abstraction of their performance, namely a contraction metric, contraction rate, and associated contraction region, which compactly characterize properties and robustness of composition. 	
Contraction metrics, which guide online control, might be learned offline through drawing on experience, or through evolution, enabling application in systems beyond the limitations of current control synthesis tools. Experiments in learning stable attractors from demonstration \citep{Billard11} can be cast as convex problems through a contraction viewpoint \citep{Ravichandar15}. This suggests that a notion of motor stability resembling contraction could guide a form of sensorimotor learning with favorable convergence.



These burgeoning extensions of contraction analysis offer an opportunity to understand and extend seemingly-complex robot control frameworks. The main contributions of this paper are to provide an analysis of Dynamic Movement Primitives (DMPs) within the framework of contraction and to introduce a new functional tool for DMPs through spatially-sparse inhibition.
Contraction analysis of DMPs provides new results related to scaling primitives in space through general diffeomorphisms, on the stability of rhythmic DMPs in general networked combinations, and robustness to parameter heterogeneity in coupled oscillators. Aside from using low-dimensional inputs to shape rhythmic high-dimensional behavior, we show that DMPs can be globally shaped through spatially-sparse modification to the DMP vector fields. This extension, which we call sparsely-inhibited DMPs (SI-DMPs) is used to manage start/stop transitions for phase oscillators in locomotion experiments with the MIT Cheetah robot. 

The paper is organized as follows. Section 2 presents DMPs and draws on commonality across varied implementations in the literature. Section 3 provides preliminaries on contraction analysis, which are then used to analyze the stability of DMPs. Section 4 builds on this analysis with an extension to sparsely inhibit Rhythmic DMPs. Section 5 presents the validation of these results to inhibit oscillations that drive locomotion in a walking gait for the MIT Cheetah robot. A short discussion and concluding remarks are provided in Section 6.

\section{Dynamic Movement Primitives}
\label{sec:DMPs}

Dynamic movement primitives \citep{Schaal12} are systems of ordinary differential equations which can be used to generate target kinematic behaviors for robotic systems. While there are many implementations of DMPs within the literature, a single DMP (i.e. not coupled to any others) is generally structured as a hierarchy of three separate systems: a reference system, canonical system, and transformation system \citep{Schaal12}. We begin by providing examples of these systems in the literature, and then describe their common general properties.

\subsection{Discrete (Point-To-Point) Motion Primitives}

	Discrete DMPs encode point-to-point motions, shaping both the behavior of the kinematic targets, as well as transients along the approach.  Letting $g$ represent a goal configuration, the state $(y,\dot{y},x) \in \real^3$ of a point-to-point DMP may be chosen to evolve as \citep{Schaal12} 
\begin{align}
\tau \ddot{y} &= k (g - y) - b \dot{y} + f(x) \label{eq:p2p_trans}\\
\tau \dot{x} &= -\alpha_x x \label{eq:p2p_phase}
\end{align}
where $k\in \real^+$, $b\in \real^+$ provide spring and damper values for a desired attractor towards the goal $g$, $\tau \in \real^+$ a temporal scaling factor and $f(x)$ a forcing function. The variables $(y,\dot{y})$ encode a position and velocity for the output of the DMP, while $x$ is a phasing variable which smoothly decays to zero. The forcing function $f(x)$ can shape the transient behavior through phased-based forcing through Gaussian basis functions
\begin{equation}
f(x) = \frac{\sum_i \Phi_i(x) w_i}{\sum_i \Phi_i(x)} x, ~~\Phi_i(x) = {\rm exp}\left({-\frac{(x - c_i)^2}{2 \sigma_1^2}  }\right)
\end{equation}
It is common to learn weights $w_i$ for these forcing functions through demonstration \citep{Schaal12}, with learning accomplished through least-squares methods. In order to increase smoothness of the output, reference systems may be employed to filter external commands, for instance with an externally provided goal $g_{ext}(t)$
\begin{equation}
\dot{g} = \alpha_g  (g_{ext}(t) - g )\,.
\label{eq:p2p_filter}
\end{equation}
Beyond translating the goal, adjustable attractor landscapes through spatial and time-based scaling have been sought as key characteristics within implementations of DMPs \citep{Schaal12}.

Consistent with the literature \citep{Schaal12} \eqref{eq:p2p_trans} is called a {\em transformation system} while \eqref{eq:p2p_phase} is called a {\em canonical system}. The role of the canonical system is to provide a notion of phase, while the transformation system uses the phase to shape the attractor landscape. Rhythmic primitives generalize this framework through the inscription of oscillations into the canonical system. 

\subsection{Rhythmic Motion Primitives}

 Letting $\x = (x_1, x_2) \in \real^2$, represent a new canonical system state, a choice for rhythmic DMP dynamics is 
\begin{align}
\tau \ddot{y} &= k (g - y) - b \dot{y} + f(\x)\\
\tau \dot{x}_1 &= \phantom{-} \omega x_2 + \rho ( r^2 - x_1^2 - x_2^2) x_1 \label{eq:hopf1} \\
\tau \dot{x}_2 &= -\omega x_1 + \rho ( r^2 - x_1^2 - x_2^2)  x_2 \label{eq:hopf2}
\end{align}
The $\dot{\x}=\f_{\x}(\x)$ dynamics in \eqref{eq:hopf1}-\eqref{eq:hopf2} are a stable Andronov-Hopf oscillator at radius $r$.\footnote{This definition differs slightly from previous canonical systems in polar coordinates $(r,\theta)$ \citep{Schaal12}. A stable limit cycle for $\x$ simplifies analysis for rhythmic DMPs here.} The forcing function $f(\x)$ provides phase-dependent forcing through von Mises bases
\begin{align}
 f(\x) = \frac{\sum_i \Phi_i(\theta(\x)) \w_i^T}{\sum_i \Phi_i(\theta(\x))} \x,~\Phi_i(\theta) = {\rm exp}\left({\frac{\textrm{cos}( \theta - \theta_i)-1}{2 \sigma_1^2}  }\right)\,
 \nonumber
\end{align}
where the angle of $\x$ denoted $\theta(\x) = \textrm{atan2}(x_2,x_1)$. Filters similar to \eqref{eq:p2p_filter} may be added to  smoothly shape references, such as the nominal center of oscillation $g$ or the oscillation amplitude $r$, in response to changes in external reference.

\subsection{Commonalities}

Across these examples, and across the literature, there is a great deal of commonality in the varied implementations of DMPs. As highlighted previously, we can typically decompose each DMP into three separate subsystems:
\begin{align}
\dot{\r} &= \f_\r(\r, \r_{ext}) & &\textrm{(Reference System)} \label{eq:common1}\\
\dot{\x} &= \f_\x(\x, \r) & &\textrm{(Canonical System)} \label{eq:common2}\\
\dot{\y} &= \f_\y(\x,\y,\r) & &\textrm{(Transformation System)} \label{eq:common3}
\end{align}
where $\r \in \mathbb{R}^{n_r}$ the reference state, $\r_{ext} \in \mathbb{R}^{n_r}$ an external command, $\x\in \mathbb{R}^{n_x}$ the canonical (phase) state, and $\y \in \mathbb{R}^{n_y}$ the transformed output. Within the categorizations provided by contraction theory, reference systems are contracting in $\r$, canonical systems are transversely contracting in $\x$, and transformation systems are contracting in $\y$. The next section provides more precise definitions of these terms and details the implications for architecting complex networks of DMPs.

\section{Contraction Analysis of DMPs}

\subsection{Contraction Preliminaries}

\label{sec:prelim}
Consider an system with state $\x\in \mathbb R^n$ and dynamics
\begin{equation}\label{eq:sys}
\dot \x = \f(t,\x)\,.
\end{equation}
Given an initial condition $\x$ at time $t=0$, $\x(t)$ denotes the flow along (\ref{eq:sys}) for $t$ seconds. We define
\begin{equation}
\A(t,\x) = \left.\pd[\f]{\x} \right|_{\x(t)}
\end{equation}
with its symmetric part $\A_s = \half \P{\A+\A\T}$.  For a symmetric matrix $\Q \in \real^{n\times n}$ we define its eigenvalues in non-increasing order $\lambda_1(\Q) \ge \lambda_2(\Q) \ge \cdots \ge \lambda_n(\Q)$. We note that $\A(t,\x )$ defines a linear time-varying system on virtual displacements $\delt{x}$ around $\x(t)$ according to $\delt{\dot{\x}} = \A(t,\x) \delt{x}$.

\begin{definition}
\citep{Slotine98} A system is said to be contracting in a forward invariant region $\calC$ if any two solutions of (\ref{eq:sys}) from different initial conditions converge to one another exponentially. Contraction can be characterized by the existence of a symmetric, uniformly positive definite metric $\M(t,\x) : \mathbb{R}  \times \calC \rightarrow \mathbb{R}^{n \times n}$ and a contraction rate $\lambda>0$, such that 
\[
\dot{\M} + \A\T\, \M + \M\, \A \le -2 \lambda \M
\]
for all $t\in \mathbb{R}$ and $\x \in \calC$.
\end{definition}

Contraction metrics provide a differential change of variables for the differential dynamics. Given a metric $\M(t,\x)$, a smooth factorization of $\M(t,\x) = \Thetab\T(t,\x) \Thetab(t,\x)$ with $\Thetab(t,\x) \in \real^{n \times n}$ provides a differential change of basis
\[
\delt{z}(t) = \Thetab(t,\x) \, \delt{x}(t) \,.
\]
Contraction conditions in $\delt{x}$ coordinates
\begin{align}
\ddt \delt{x}\T \M \delt{x} &= \delt{x}\T \left( \dot{\M} + \A\T \M + \M\A \right) \delt{x} \\ &\le -2 \lambda \delt{x}\T \M \delt{x}
\end{align}
are equivalent to the following in $\delt{z}$:
\begin{align}
\ddt \delt{z}\T \delt{z} = 2\delt{z}\T \F_s \delt{z} \le -2\lambda \delt{z}\T \delt{z} 
\end{align}
where $\F = \left(\Thetab \A + \dot{\Thetab}\right)\Thetab^{-1}$ is called a generalized Jacobian of $\A$ associated with the differential change of coordinates $\Thetab$. Thus, the contraction conditions are equivalent to the existence of a differential change of coordinates $\Thetab$ such that $\lambda_1(\F_s) \le -\lambda$. As a matter of convention, contraction rates $\lambda$ will be expressed as positive numbers, and the eigenvalues of the associated generalized Jacobian uniformly negative.

All of the above results apply to the use of the Euclidean norm to characterize convergence. This can be generalized \citep{Slotine98}. Take any norm ${|\cdot|:\mathbb{R}^n \rightarrow \mathbb{R}}$, with its induced norm denoted $\|\cdot\|$. The associated matrix measure $\mu$ is defined as $\mu(\A)= \lim_{h\rightarrow0^+} \tfrac{1}{h}(\|\I + h \A\| - 1)$, originally introduced in \citep{Lozinskii59,Dahlquist59}. See \citep{Vidyasagar} for a more current treatment and \citep{Desoer72} for relevant early applications. Under the Euclidean norm, $\lambda_1(\F_s) \le -\lambda$ is equivalent to $\mu(\F) \le -\lambda$. More generally a system is contracting if there exists a matrix measure such that $\mu(\F)\le-\lambda$. It is important to emphasize that the freedom in norm is separate from and in addition to the freedom in metric when it comes to obtaining contraction certificates. Throughout the manuscript, unless otherwise specified, the Euclidean norm is assumed.

For systems which possess orbits, such as Rhythmic DMPs, perturbations in phase are persistent in time and thus cannot be contracting. However, relaxing contraction along the flow the of the system provides a useful related property of Transverse Contraction.

\begin{definition}
\citep{Manchester14} An autonomous system is said to be transverse contracting in a compact, strictly forward invariant region $\calK$ if any two solutions of (\ref{eq:sys}) from different initial conditions converge to one another exponentially up to a monotonic reparameterization of time. Transverse contraction is characterized by the existence of a time-invariant symmetric, uniformly positive definite metric $\M(\x) : \calK \rightarrow \mathbb{R}^{n \times n}$ and a contraction rate $\lambda>0$. Such that
\begin{equation}
\delt{x}^T \left(\dot{\M} + \A\T\, \M + \M\, \A  + +2 \lambda \M \right) \delt{x}  \le 0  \label{eq:trans_condition}
\end{equation}
for all $\x \in \calK$ and for all $\delt{x} \ne 0$ with $\f(\x)^T \M(\x) \delt{x} = 0$.
\end{definition}
Intuitively, \eqref{eq:trans_condition} relaxes the contraction condition along the vector field $\f(\x)$ by enforcing that only displacements transverse to the flow need be contracting. A main implication of a system being transverse contracting applies when the region $\calK$ does not have an equilibrium.

\begin{prop} \citep{Manchester14}
If $\f(\x)\ne0$ for all $\x \in \calK$ and $\f$ transverse contracting on $\calK$, then the solution to \eqref{eq:sys} from any initial condition in $\calK$ approaches a unique limit cycle.
\label{prop:limitCycle}
\end{prop}

\begin{theorem}
Suppose the system (\ref{eq:sys}) is autonomous and has a compact transverse contraction region $\calK$. Then there exists a differential change of coordinates $\delt{y} = \Thetab(x)  \delt{x}$ such that its generalized Jacobian $\F$ satisfies $\lambda_1(\F_s) =0$ and $\lambda_2(\F_s)<0$  uniformly. 
\end{theorem}
\begin{pf}
See Appendix A. 
\end{pf}

\subsection{Scaling in Space and Time}
\label{sec:spaceandtime}
A central requirement of DMPs is an ability to scale primitives in space and time \citep{Schaal12}. A contraction viewpoint readily provides a new result towards a general class of system scaling operations. Assume a transformation system $\dot{\y} = \f_\y(\y,\x,\,\r)$ contracting in $\y$ under metric $\M(\y,\x,\r)$, and a smooth diffeomorphism $\y' = \TT(\y)$. Letting $\J = \pd[\TT]{\y}$, time-scaled dynamics for $\y'$ can be formed to follow 
\begin{equation}
\tau(t)\, \dot{\y}' = \J(\y)\, \f_\y( \y, \x,\r) |_{\y = \TT^{-1}(\y')}
\label{eq:differ}
\end{equation}
with $\tau(t)>0$ uniformly. This system is contracting in $\y'$ under metric $\M' = \J^{-T} \M \J^{-1}$. An analogous result holds for a diffeomorphism applied to a transverse contracting system. Scaled primitives in time and in space have been pursued to shape transformation systems in $\mathbb{R}^2$ and $\mathbb{R}^3$ \citep{Schaal12}. The above result suggests this approach can be employed more broadly to shape DMP dynamics on $\mathbb{R}^n$. For instance, given a an Andronov-Hopf oscillator in $\mathbb{R}^2$ with appended state dynamics $\dot{x}_3= -x_3, \ldots,\dot{x}_n = -x_n$, this transverse contracting system in $\mathbb{R}^n$ could be sought to provide a target canonical limit cycle in an $n$-DoF robot arm through design of a diffeomorphism $\TT$.

A useful special case of the above result pertains to homogeneous transformations with scaling. When $\TT(\y) = s \R \y +\y_{\TT}$, for $\R \in \mathsf{SO}(n_y)$, $s\in\mathbb{R}^+$, $\y_{\TT}\in\mathbb{R}^{n_y}$, the entire attractor landscape for $\y$ undergoes rotation, scaling, and translation when applied to $\y'$. Thus, contracting systems can be viewed in a sense as mother systems, akin to wavelets, with scaling in space and time providing contracting daughter systems. Additive copies of different daughter systems could be be sought, similar to the linear combinations of primitive attractors found in frogs \citep{Bizzi94,Slotine01}. In this light, a contraction viewpoint may also allow primitives to be used for multi-scale approximations of attractor dynamics, providing bases for the coarse and fine grains of motion in a precise theoretical context.  The optimization of such multi-scale transformations, as opposed to learning underlying contracting dynamics themselves, presents a new area for study in DMP learning. 

\subsection{Combination Properties}
Contracting systems possess useful compositional properties, retaining contraction through system combinations such as parallel interconnections, hierarchies, and certain classes of negative feedback \citep{Slotine98}. Combinations of transverse contracting and contracting systems enjoy similar properties in certain cases (See \cite{Manchester14} for details). We state three results which will simplify the stability analysis of DMPs.

\begin{prop} \citep{Slotine98} 
If $\f_1(t,\x_1)$ contracting, and $\f_2(t,\x_2,\x_1)$ contracting for each fixed $\x_1$, then the hierarchy $\dot{\x}_1 = \f(t,\x_1)$, $\dot{\x}_2 = \f(t,\x_2,\x_1)$ is contracting.
\label{prop:contrheirarchy}
\end{prop}

\begin{prop} \citep{Manchester14}
If $\f_1(\x_1)$ contracting, and $\f_2(\x_2,\x_1)$ transverse contracting for each fixed $\x_1$, then the hierarchy $\dot{\x}_1 = \f(\x_1)$, $\dot{\x}_2 = \f(\x_2,\x_1)$ is transversely contracting.\label{prop:tcontr1}
\end{prop}

\begin{prop} \citep{Manchester14}
If $\f_1(\x_1)$ transverse contracting, and $\f_2(\x_2,\x_1)$ contracting for each fixed $\x_1$, then the hierarchy $\dot{\x}_1 = \f(\x_1)$, $\dot{\x}_2 = \f(\x_2,\x_1)$ is transversely contracting.
\label{prop:tcontr2}
\end{prop}

\subsection{Contraction Analysis of DMPs}

Discrete DMPs such as \eqref{eq:p2p_trans}-\eqref{eq:p2p_phase} employ a canonical system that is exponentially stable -- and thus contracting. We introduce the following generalization. 

\begin{thm} Assume a discrete DMP wherein \eqref{eq:common1} is contracting in $\r$, \eqref{eq:common2} contracting in $\x$, and \eqref{eq:common3} contracting in $\y$. Then the overall hierarchy \eqref{eq:common1}-\eqref{eq:common3} is contracting. 
\label{thm:DiscrteDMP}
\end{thm}

\begin{figure}
\center
\includegraphics[width = .6\columnwidth]{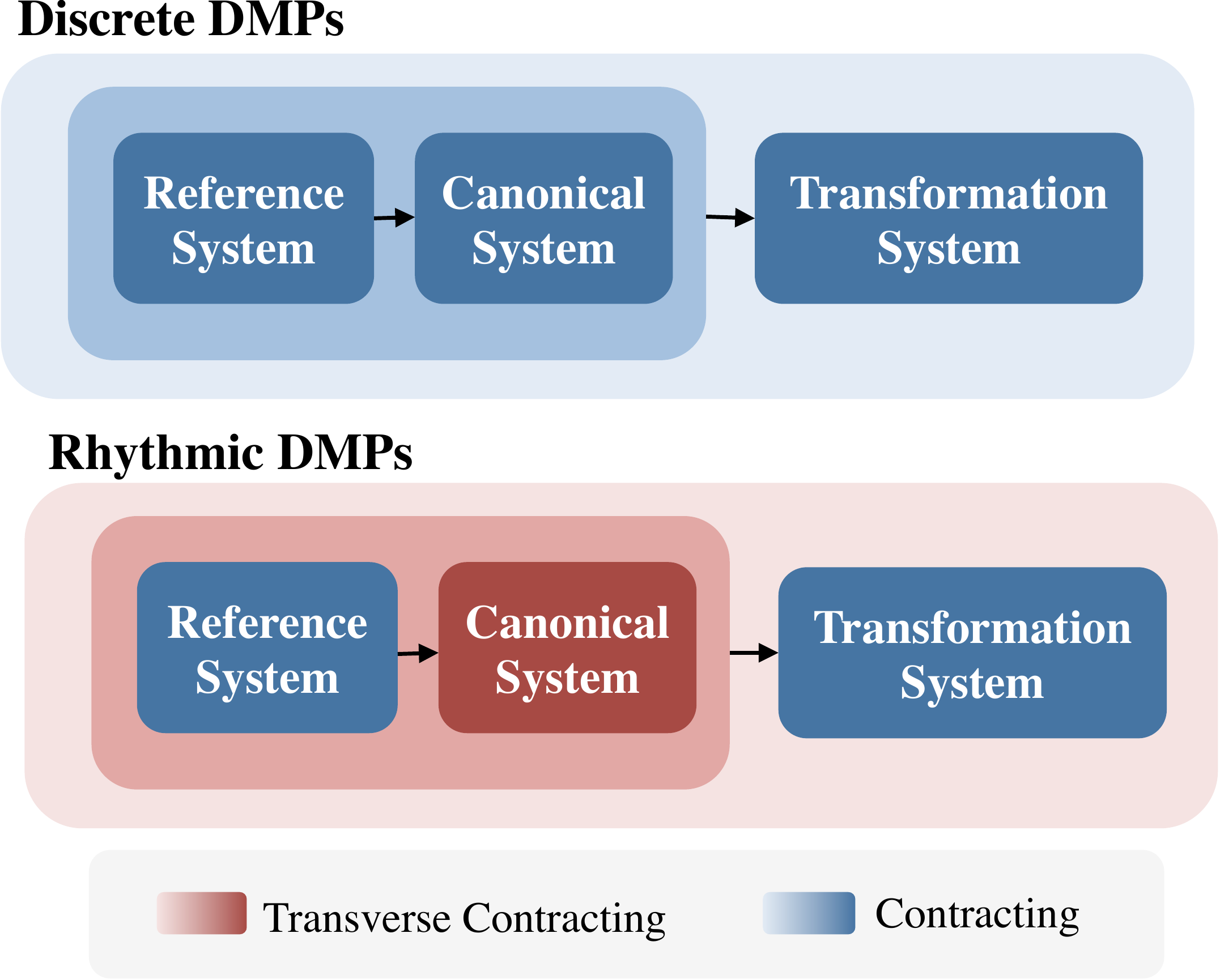}
\caption{Contraction analysis readily admits stability specifications for discrete and rhythmic DMPs.}
\label{fig:ContractionOfDMPs}
\end{figure}

\begin{pf}
In the spirit of \cite{Slotine06c}. Applying Proposition \ref{prop:contrheirarchy} to \eqref{eq:common1} in hierarchy with \eqref{eq:common2} shows that \eqref{eq:common1}-\eqref{eq:common2} is contracting jointly in $\r,\x$. Repeated application of this system in hierarchy with \eqref{eq:common3} provides the desired result. This is sketched in Figure \ref{fig:ContractionOfDMPs}. \hfill$\qed$
\end{pf}

A similar result holds in the case of rhythmic DMPs, wherein a transverse contracting canonical system percolates the transverse contraction property to the rhythmic DMP as a whole. Its proof follows Thm.~\ref{thm:DiscrteDMP}, except using Propositions~\ref{prop:tcontr1} and \ref{prop:tcontr2}. This result is depicted in Figure \ref{fig:ContractionOfDMPs}.

\begin{thm} Assume a rhythmic DMP wherein \eqref{eq:common1} is contracting in $\r$, \eqref{eq:common2} transverse contracting in $\x$, and \eqref{eq:common3} contracting in $\y$. Then, for a fixed external reference $\r_{ext}$ the overall hierarchy \eqref{eq:common1}-\eqref{eq:common3} is transverse contracting. 
\label{thm:RhythmicDMPs}
\end{thm}

	In the case of $N$ coupled DMPs (rhythmic or discrete), assume a single reference vector $\r$, with canonical states $\x = \{ \x_1,\,\ldots,\,\x_N\}$, and transformation states $\y = \{ \y_1,\,\ldots,\,\y_N\}$. Theorems \ref{thm:DiscrteDMP} and \ref{thm:RhythmicDMPs} can be used to assert contraction for the coupled attractors. We discuss the case of CPGs to illustrate the application of this result.

\begin{figure}
\center
\includegraphics[width= .35\columnwidth]{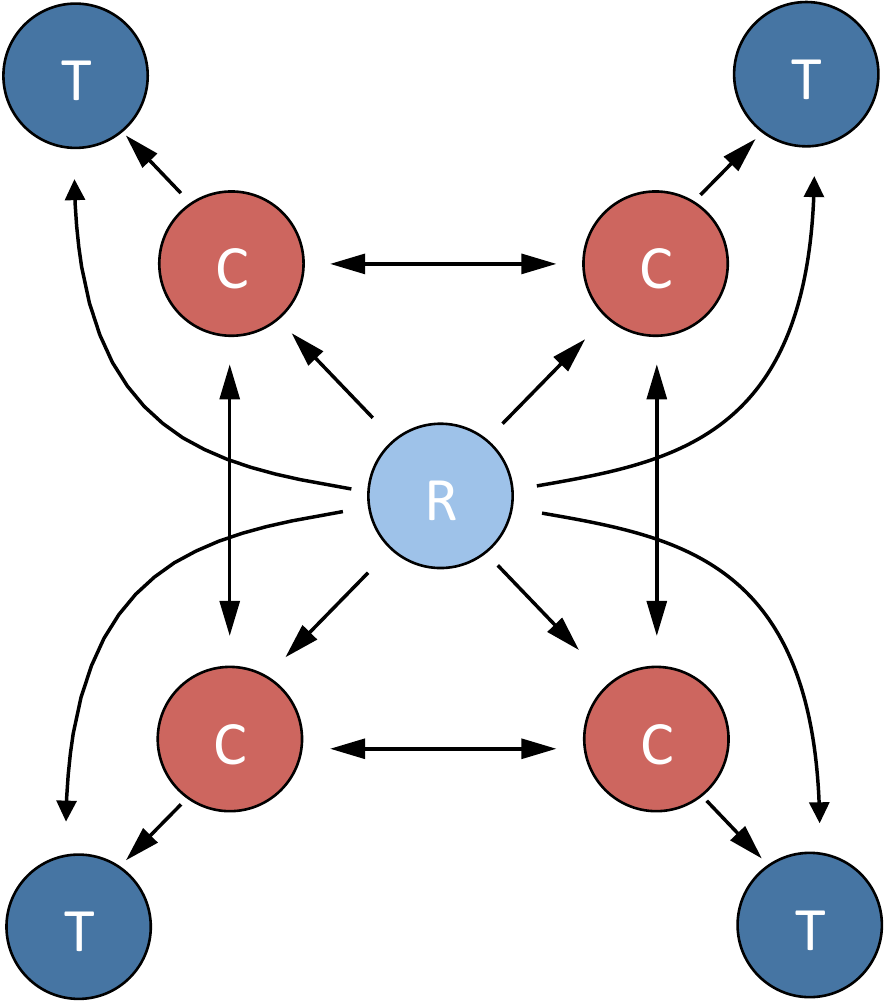}
\caption{Network of coupled rhythmic DMPs with common reference system (R), coupled canonical systems (C), and transformation systems (T).}
\label{fig:coupledDMPS}
\end{figure}

	CPGs can be interpreted to represent a network of rhythmic DMPs with coupling exclusively through phase variables $\x$. Assuming a common reference vector $\r$ for $N$ DMPs, as shown in Fig.~\ref{fig:coupledDMPS} for $N=4$, coupled diffusively through their phase variables $\x_1,\ldots,\x_N$. Assume further that coupling occurs through neighbors $\mathcal{N}_i$ according to:	
\begin{align}
\dot{\r}       &= \f_\r(\r,\r_{ext}) \label{eq:CPGnetwork1}\\[.5ex]
\dot{\x}_i &= \f_\x(\x_i,\r) + \sum_{j\in \mathcal{N}_i} \K_{ij} (\x_j - \x_i)
\label{eq:coupledDMPs} \\
\dot{\y}_i &= \f_{\y i}(\y_i, \x_i, \r) \label{eq:CPGnetwork3}
\end{align} 
for some set of gains matrices with each $\K_{ij}= \K_{ji}$ and $(\K_{ij})_s>0$\,. When $\f_\x$ is an Andronov-Hopf oscillator as in \eqref{eq:hopf1}-\eqref{eq:hopf2} with gains $\K_{ij} = k \mathbf{I}$, the canonical systems are guaranteed to asymptotically synchronize \citep{Chung10b} (i.e. $\x_1=\cdots=\x_N$). Combining synchronization results from \cite{Slotine05} with contraction results from \cite{Manchester14} allows this result to be generalized.

\begin{thm} Assume a network of $N$ rhythmic DMPs \eqref{eq:CPGnetwork1}-\eqref{eq:CPGnetwork3} whose individual uncoupled dynamics $(\f_\r, \f_\x, \f_{\y i})$ satisfy the assumptions of Thm.~\ref{thm:RhythmicDMPs}. Let $\A_i=\left.\partial\f_{\x}/\partial\x\right|_{\x_i}$ and $\L_\K$ the symmetric part of the weighted block-Laplacian matrix \citep{Slotine05} from the graph $\mathcal{G}$ with edges $\cup_{i} \{i\} \times \mathcal{N}_i$. If $\mathcal{G}$ is connected and
\begin{equation}
\lambda_{N+1}(\L_\K) > {\max}_i \lambda_{\rm{max}}(\A_{is})
\label{eq:eigcond}
\end{equation}
uniformly, then \eqref{eq:CPGnetwork1}-\eqref{eq:CPGnetwork3} is transverse contracting. \label{thm:CoupledRhythmicDMPs}
\end{thm}
\begin{pf}
Graph connectivity and \eqref{eq:eigcond} guarantee asymptotic synchronization of  canonical states \citep{Slotine05}. Transverse contraction of the {\em reduced system} $\dot{\z} =\f_\x(\z,\r)$ (to which each $\x_i$ converge) implies transverse contraction of the coupled systems. Thm.~\ref{thm:RhythmicDMPs} then ensures transverse contraction for the coupled DMPs.\hfill$\qed$
\end{pf}


\begin{rem} Note that the requirement of a common reference system in Thm.~\ref{thm:CoupledRhythmicDMPs} satisfies  input-equivalence conditions from previous synchronization studies \citep{Pham07}. The results on combination properties  from this previous work could be pursued to analyze couplings between yet other modules in the DMP network.  
\end{rem}

\begin{rem}
Suppose $\V \in\mathbb{R}^{n (N-1) \times nN}$ with orthonormal rows, such that its nullspace represents the synchronization subspace for the $\x$ dynamics. Then \eqref{eq:eigcond} can be phrased equivalently as $\max_i\mu(\A_i) + \mu(-\V \L_\K \V^T) <0 $. This follows from \citep[Theorem 3]{Russo11} and the fact that $\mu(\A+{\mathbf{B}})\le\mu(\A)+\mu({\mathbf{B}})$ for any matrix measure. Recent work \citep{Leonard16} has shown that, in comparison to a Euclidean contraction analysis, nonsmooth Lyapunov analysis can achieve tighter critical coupling strength bounds within certain parameter ranges for coupled neural oscillator models. This suggests that a practitioner may consider the conditions $\max_i\mu(\A_i) + \mu(-\V \L_\K \V^T) < 0$ under different norms to limit required coupling gains. See also \citep{Sontag13} for a more general discussion on matrix measures for contraction analysis of networked systems.
\end{rem}

\newcommand{\bomega}{\boldsymbol{\omega}}
\subsection{Coupled Oscillators with Multiple Frequencies}
Contraction analysis also sheds light onto the case when heterogeneous canonical oscillators with multiple frequencies are coupled in networked combinations. When coupling systems to the physical world, natural passive dynamics of compliant mechanisms \citep{Williamson99} or low-level control loops \citep{Chung10} might be fixed. The coupling of these systems with CPG oscillators requires reasoning about coupled heterogeneous oscillators. Despite empirical observations on the robustness of such couplings to heterogeneity \citep{Chung10}, analytical results are largely lacking. We provide a brief discussion below which shows the capability of tools from transverse contraction to describe these phenomena. The implications of these results extend beyond robotics, and e.g., may illuminate entrainment mechanisms when driving spiking neurons, as in \citep{Mainen95}.

Assume that the feedback-coupled oscillators \eqref{eq:coupledDMPs} are not identical, but instead are each parameterized continuously by parameters ${\boldsymbol \omega_i}\in \mathcal{P}$.
\begin{equation}
\dot{\x}_i = \f_\x(\x_i, \r,{\boldsymbol \omega}_i) + \sum_{j\in \mathcal{N}_i} \K_{ij} (\x_j - \x_i)
\label{eq:hetero}
\end{equation}
It is assumed that each uncoupled system $\dot{\x}_i =\f_\x(\x_i, \r,{\boldsymbol \omega}_i)$ is transverse contracting for $\bomega_i \in \mathcal{P}$.
\begin{prop}
Assume a nominal parameter selection ${\boldsymbol \omega}_0 \in \rm{int}(\mathcal{P})$ such that, when each $\bomega_i = {\bomega_0}$, the coupled canonical systems \eqref{eq:hetero} are transverse contracting with rate $\lambda>0$ under a metric $\M(\x)$ in a region $\calK$ with no equilibria. Then, there exists an open set $\mathcal{W} \subset \mathcal{P}$ such that $\bomega_0 \in \mathcal{W}$ and, if each $\bomega_i \in \mathcal{W}$ then the coupled heterogeneous oscillators \eqref{eq:hetero} are transverse contracting on $\calK$ under $\M(\x)$. The coupled system asymptotically approaches a unique limit cycle $\mathcal{O}$ with period $T>0$.
\end{prop}
\begin{pf}
Transverse contraction is a topologically open condition, with transverse contraction rate $\lambda > 0$ uniformly on the compact strictly forward invariant region $\mathcal{K}$. The condition that the coupled oscillators with $\bomega_i = \bomega_0$ have no equilibrium on $\calK$ is also an open condition. Thus, if $\f_\x(\x_i, \r, \bomega_i)$ depends continuously on $\bomega_i$, there is an open set $\mathcal{W}$ containing $\bomega_0$ such that when each $\bomega_i \in \mathcal{W}$, 1) $\mathcal{K}$ remains forward invariant, 2) transverse contraction conditions under $\M(\x)$ hold with rate $\epsilon \lambda$ for some $\epsilon >0$, and 3) the coupled heterogeneous oscillators have no equilibrium in $\mathcal{K}$. When each $\bomega_i \in \mathcal{W}$, Proposition \ref{prop:limitCycle} guarantees a unique limit cycle $\mathcal{O}$ with common period $T>0$. \hfill$\qed$
\end{pf}
Intuitively, this result is reminiscent of how contraction at a point can be extended to contraction within a guaranteed basin of attraction \citep{Slotine98}.

Note that when each $\bomega_i \in \mathcal{W}$, each $\x_i$ is bounded due to forward invariance of $\cal{K}$. Thus, the mismatch $\mathbf{d}_i = \f_\x(\x_i, \r, \bomega_i)-\f_\x(\x_i, \r, \bomega_0)$ remains bounded. Viewing the heterogeneous oscillators with $\bomega_i \ne \bomega_0$ as a disturbance on the case when each $\bomega_i = \bomega_0$,
\begin{equation}
\dot{\x}_i = \f_\x(\x_i, \r,{\boldsymbol \omega}_0) + \sum_{j\in \mathcal{N}_i} \K_{ij} (\x_j - \x_i) + \mathbf{d}_i
\end{equation} 
Let $\mathbf{d} = \{\mathbf{d}_1,\ldots,\mathbf{d}_N\}$ collect the disturbances and suppose $\sup_t |\mathbf{d}(t)| = \overline{d}$. Robustness results from \cite{Slotine05} guarantee the existence of $r>0$ (dependent on $\M$ alone) such that all $| \x_j - \x_i | \le \frac{r}{\lambda}\overline{d}$ after exponential transient. This implies that as gains $\K_{ij}$ are increased, synchronization errors 
can be made arbitrarily small. 

\begin{figure}
\center
\includegraphics[width = .95\columnwidth]{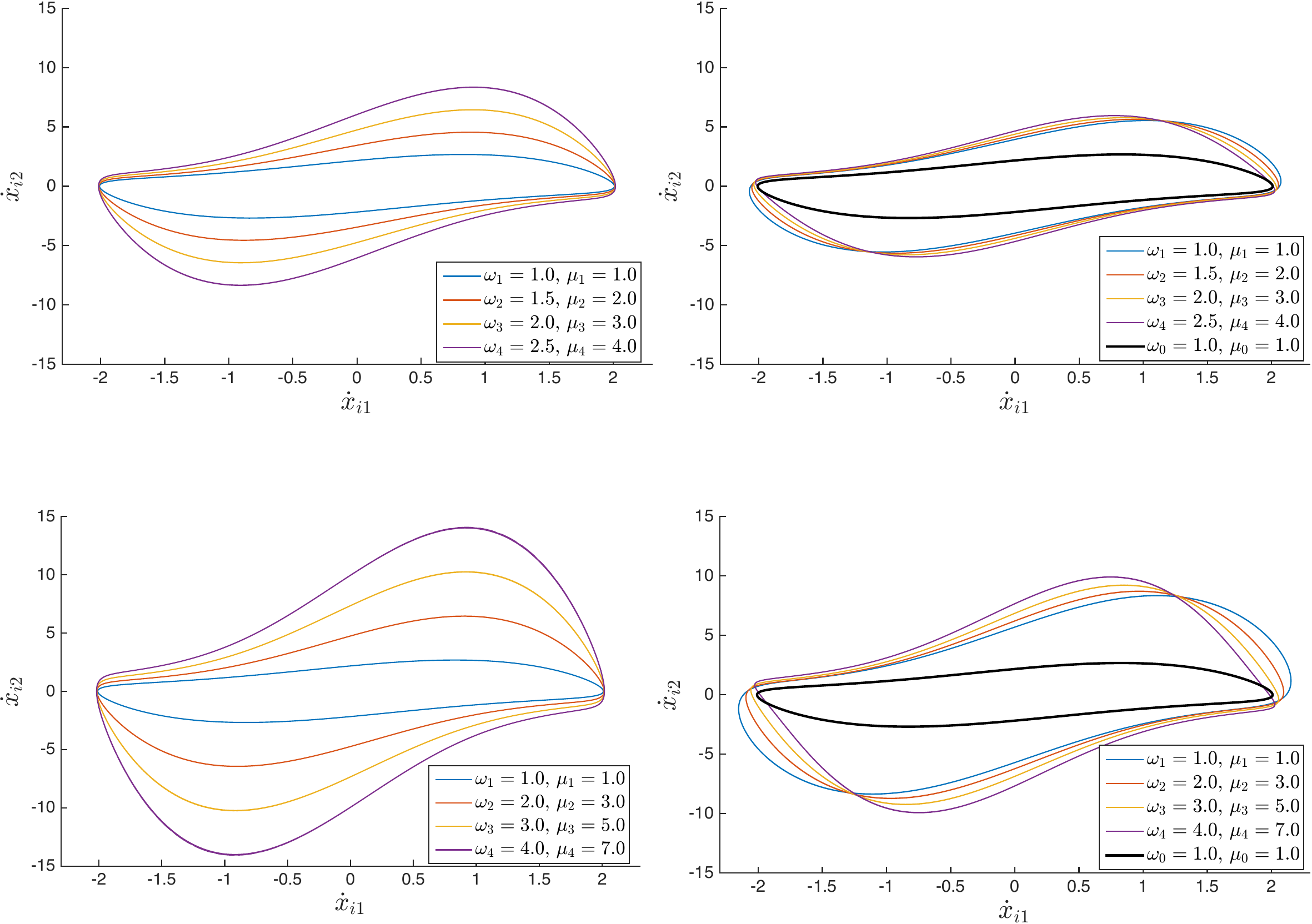}
\caption{Left: Uncoupled heterogeneous Van der Pol oscillators with $\dot{x}_{i1} = x_{i2}$ and $\dot{x}_{i2} = -\omega_i^2 x_{i1} + \mu_i ( 1 - x_{i1} ) x_{i2}$. Right: With diffusive coupling $\K_{ij} = \rm{diag}(4,4)$. New results guarantee a common period for changes in parameter heterogeneity within an open set, and continuity of the resulting periodic orbit across parameters.} 
\label{fig:vdp}
\end{figure}

Transverse contraction analysis allows for us to further assert a region where the limit cycle $\mathcal{O}$ must reside. Assume a transverse contracting system with rate $\lambda$ subject to disturbance $\mathbf{d}$. It is straightforward to show, using Euler-Lagrange conditions on the geodesics underlying $\M(\x)$ \citep{Singh17},  that any perturbed trajectory stays within a tube of radius $\frac{R}{\lambda} \overline{d}$ around its unperturbed trajectory. Again, $R>0$ depends on $\M$ alone.  This result is stated formally and proved in Appendix \ref{sec:disturbed}.  Thus, for parameters near the homogeneous parameter set, the limit cycle $\mathcal{O}$ for the heterogeneous oscillators varies continuously. Figure \ref{fig:vdp} shows an example of coupling heterogeneous Van der Pol oscillators. Coupled oscillators reach a common period despite significant  heterogeneity.

\begin{rem}
It is interesting to note that transverse contraction on a simply connected $\calK$ implies contraction on $\calK$ \citep{Manchester14e} due to results from \citep[Thm.~3.1]{Leonov96} and \citep[Thm.~4]{Manchester14}. Thus, for parameter changes beyond $\mathcal{W}$, the coupled system may remain transverse contracting but on a new, and simply connected region. In this case, parameter differences result in contracting behavior. Thus, since the system is autonomous, it will tend  towards a unique equilibrium.
\end{rem}

\section{Sparsely-Inhibited Rhythmic DMPs}

We have seen that DMPs allow a sparse set of reference inputs $\r_{ext}$ to effectively shape the high-dimensional attractor landscape of discrete and rhythmic DMPs. This sections builds towards the ability switch between rhythmic and discrete DMPs through only spatially-sparse modification to the vector fields of the canonical system. 

\subsection{Local Influence of Contracting Dynamics}

Here we show that if a transverse contraction region contains a contraction region, then all trajectories tend to a unique equilibrium.  This will be a motivating mechanism in sparse control of transverse contraction.

\begin{thm}
Consider an autonomous system. Let $\calK$ a transverse contraction region and $\calC$ a contraction region. Furthermore, assume that $\calK \cap \calC \ne \emptyset$. Then, within ${\calK \cup \calC}$ there is a unique equilibrium $\x^*$. Such $\x^*$ satisfies $\x^* \in {\calK\cap \calC}$ and the solution from any initial condition $\x_0 \in {\calK\cup \calC}$ satisfies $\x_0(t) \rightarrow \x^*$ exponentially as $t \rightarrow \infty$.
\label{thm:Intersection}
\end{thm}
\begin{pf}
Since $\calC$ contracting, there is a unique equilibrium $\x^*$ contained in $\calC$. Furthermore, the solution from any initial condition $\x_c \in \calC\cap \calK$ satisfies  $\x_c(t) \rightarrow \x^*$ exponentially as $t\rightarrow \infty$. Since $\calK$ strictly forward invariant, this implies that $\x^* \in \calC\cap\calK$. In addition, by $\calK$ transverse contracting, there exists a strictly monotonic reparameterization of time $\kappa(t)$ such that the solution from any initial $\x_k \in \calK$ satisfies $\x_k(t) \rightarrow \x^*( \kappa(t)) = \x^*$ exponentially.  \hfill$\qed$
\end{pf}

\begin{rem}
This theorem may be combined with contraction tools for sequential composition methods \citep{Burridge99,Tedrake10} as proposed in \citep{Slotine01}. All states in the outer contraction region $\calK$ are funneled to states in an inner contraction region $\calC$ in the above theorem.  In this spirit, a discrete set of controllers $i=0,\ldots, n_c$ which provide nested regions $\mathcal{C}_i \subset \mathcal{K}_i$ could be sought to guide controller switching. If each $\calC_i \subset \calK_{i+1}$, then for any initial condition $\x \in \calK_0$ a nesting of the above theorem ensures the existence of a switching sequence which transports $\x(t)$ to $\calC_{n_c}$. Contraction metrics may represent a more flexible alternative to Lyapunov-based characterizations of composability, as existing Lyapunov-based methods rely on explicit reference trajectories.
\end{rem}

With this theorem as motivation, we consider how the addition of a contracting vector field influences a transverse contracting system. 

\begin{prop}
Assume two vector fields $\dot{\x} = \f_1(\x)$ and $\dot{\x}=\f_2(\x)$ such that $\f_2$ renders a compact region $\calC$ contracting with rate $\lambda_2$ under a metric $\M(\x)$. Then, there exists some $\alpha_0>0$ such that for all $\alpha > \alpha_0$ the vector field $\dot{\x} = \f_1(\x) + \alpha \f_2(\x)$ is contracting on $\calC$ under $\M$.
\label{prop:strongEnoughLocalContraction}
\end{prop}

\begin{pf}
Let $\f = \f_1 + \alpha \f_2$ for $\alpha > 0$. Then
\begin{align}
\dot{\M} + \A\T \M + \M \A &\le \pd[\M]{\x}\cdot\f_1 + \A_1\T \M + \A_1 \M -2 \alpha \lambda_2 \M\,\nonumber
\end{align}
since $\f_2$ contracting under $\M$ with rate $\lambda_2$. Let
\[
\beta = \inf \left\{ b \in \mathbb{R} ~{\Large |}~ \forall{\x} \in \calC,~ \pd[\M]{\x}\cdot\f_1 + \A_1\T \M + \A_1 \M < b \M \right\}  
\]
Letting $\alpha_0 = \frac{\beta}{2 \lambda_2}$ it follows that any $\alpha > \alpha_0$ renders $\f_1+\alpha \f_2$ contracting on $\calC$ under $\M$. \hfill$\qed$
\end{pf}

\subsection{Application to Sparse Inhibition of Rhythmic DMPs}
	
	Proposition \ref{prop:strongEnoughLocalContraction} can be used to sparsely inhibit networks of rhythmic DMPs. Assume a network as \eqref{eq:CPGnetwork1}-\eqref{eq:CPGnetwork3} with coupling only through canonical variables $\x = \{\x_1,\ldots,\x_N\}$. Also assume that the network satisfies the assumptions of Thm.~\ref{thm:CoupledRhythmicDMPs}. The coupled canonical dynamics for $\x$ decompose into the sum of a nominal transversely contracting component and a semi-contracting coupling component
\[
\ddt \x = \F_\x(\x,\r) - \L\, \x
\] 
where $\F_\x = \{ \f_\x(\x_1,\r), \ldots , \f_\x(\x_N,\r)\}$ and $\L$ is the block-Laplacian matrix of the network satisfying $\L_s = \L_\K$ as defined previously. Note that $\L_\K$ is positive semi-definite quantity, and due to connectedness of the graph, $\L_\K \x  = \mathbf{0}$ iff $\x_1=\cdots=\x_N$. Assume now that an additional influence $\g(\x_1)$, contracting in the identity metric, is added to the dynamics for $\dot{\x}_1$. Then 
\begin{equation}
\ddt \x = \F_\x(\x,\r) - \L\, \x + [\g(\x_1)^T, \mathbf{0}, \ldots, \mathbf{0}]^T\
\label{eq:inhibitedCanonicalSystem}
\end{equation}
Letting $\f_{\rm inh}=  - \L\, \x + [\g(\x_1)^T, \mathbf{0}, \ldots, \mathbf{0}]^T$, examining the symmetric part of the Jacobian in $\x$ reveals:
\begin{align}
\delt{x}^T  \pd[\f_{\rm inh}]{\x}  \delt{x} = -\delt{x}^T  \L_\K \delt{x}  + \delt{x}^T_1 \left( \pd[\g]{\x_1} +  \pd[\g]{\x_1}^T \right) \delt{\x}_1
\nonumber
\end{align}
Yet, since $\g$ contracting under the identity metric, the above is negative definite. Thus $\f_{\rm inh}$ is contracting.

Intuitively, a connected network topology allows contraction for a single node to percolate to contraction for the coupled network. In this light, \eqref{eq:inhibitedCanonicalSystem} decomposes as the sum of a transverse contracting vector field with a contracting vector field. Prop.~\ref{prop:strongEnoughLocalContraction} thus ensures that for strong enough coupling gains and a strong enough influence of $\g(\x_1)$, the entire coupled transversely contracting network will transform into a contracting network through influence of $\g(\x_1)$ alone. Indeed, since the transverse contraction conditions admit a unique orbit, such influence can be activated locally, anywhere along the orbit, in order to capture the oscillations of the entire network. 

With this general networked systems view, sparse control could likely be applied to other contexts as well, for instance to modulate biochemical oscillations in the brain \citep{Mainen95,Canter16}. When oscillations follow predictable patterns, spatially sparse control allows the natural dynamics of the system to provide convergence to a desired area before expending effort. Such mechanisms could potentially offer energetic benefits in the application of inhibition in biochemical processes.

\begin{rem}
The ability to sparsely influence the qualitative behavior of large networks is also reminiscent of leader-follower networks and oscillator death through topologically-sparse network modification \citep{Slotine05}. The behavior of coupled oscillators have also shown an ability to inhibit or incite oscillations through temporally-sparse forcing \citep{Gerard06}. 
\end{rem}

\begin{rem}
The conditions of a bidirectional coupling $\K_{ij} = \K_{ji}$ can be relaxed to consider directional couplings in the case that $\K_{ij} = \K = \K^T>0$ for a fixed $\K$. The sparse inhibition result holds if all nodes are reachable from the inhibited node \citep{Caughman06}. \end{rem}


\section{Experiments With the MIT Cheetah}

\begin{figure}
\center
\includegraphics[width=.8\columnwidth]{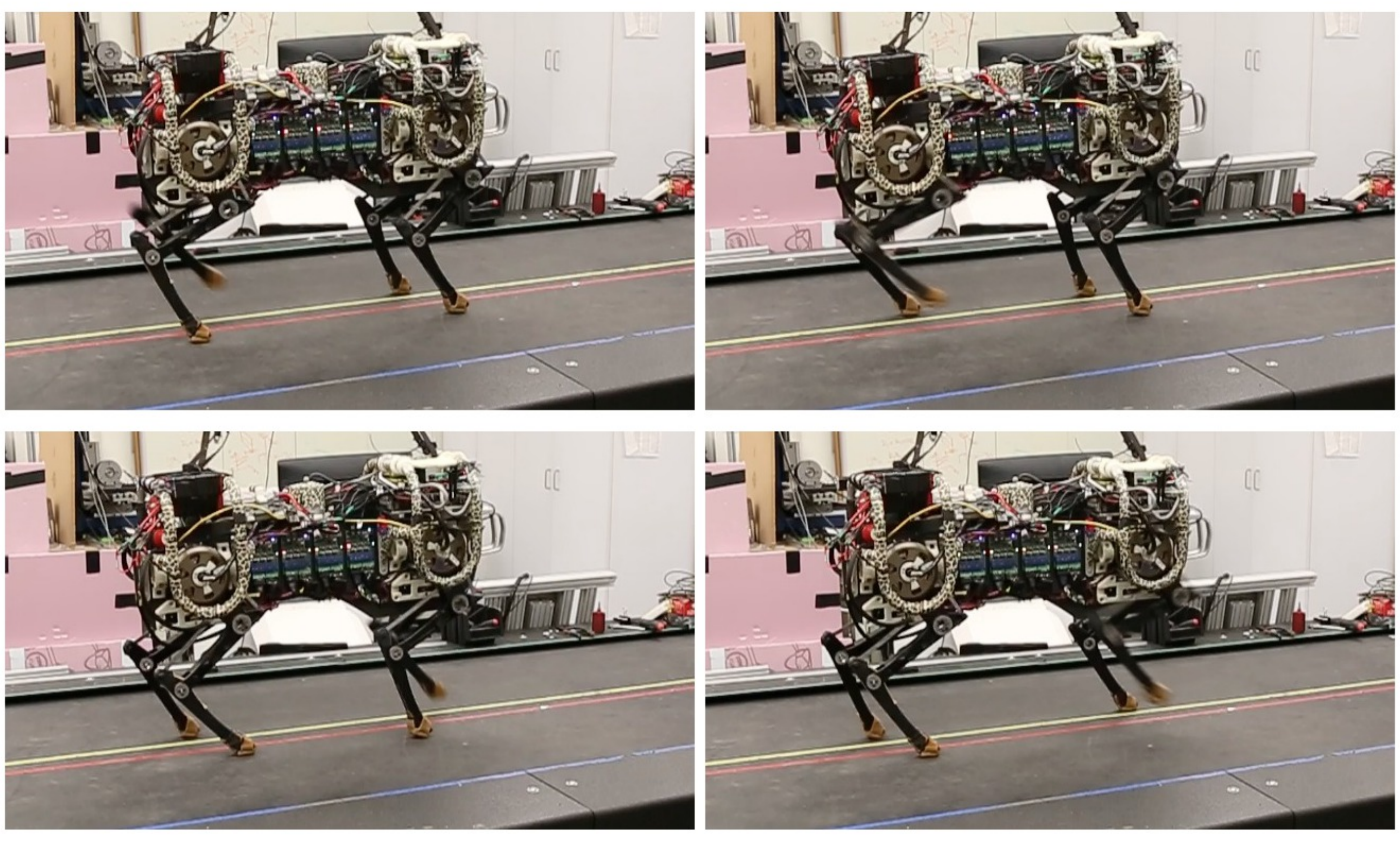}
\caption{SI-DMPs are applied to provide rhythmic oscillations for a four-beat amble gait.}
\label{fig:gait_amble}
\end{figure}

	This section describes experiments using SI-DMPs with the MIT Cheetah robot. The MIT Cheetah is a quadrupedal robot driven by brushless DC-electric motors capable of force-controlled operation with an ability to render ground reaction forces at rates of up to 100 Hz \citep{WensingUnpub15}. These actuator capabilities and simple-model-based control enable the Cheetah to bound at speeds of up to 6 m/s \citep{WensingUnpub16} and to autonomously jump over obstacles \citep{Park15b}. Moving towards a more diverse set of gaits, the use of CPGs provides a promising method to stabilize the gait pattern from step to step and to smoothly transition this pattern between gaits. We report here on the use of SI-DMPs to manage start-stop transitions in a four-beat amble gait as show in Figure \ref{fig:gait_amble}. A video is provided online at
	{\tt https://youtu.be/v4d4CrKX1k0}.

	The reference system for the application of Rhythmic DMPs consists of a desired speed $v$ and turn rate $\gamma$. Both are provided through a first-order low-pass filter
\begin{align}
\dot{v}      &= \alpha_r (v_{ext} - v) \\
\dot{\gamma} &= \alpha_r (\gamma_{ext} - \gamma)
\end{align}
Four Andronov-Hopf oscillators are used for all-to-all phase coupling with gains $\K_{ij} = k \mathbf{I}$ to synchronize the leg phases. Rotational invariance of the Andronov-Hopf dynamics admits a change of variables to encode the desired phase offset $\phi_{ij}$\footnote{This  approach holds under looser conditions that only require rotational invariance of $\f_\x$ under rotations by $\phi_{ij}$ for all $(i,j) \in \mathcal{G}$. That is, $\f_\x(\x_j,\r) = \R(\phi_{ij})^T \f(\R(\phi_{ij}) \x_j, \r)$ $\forall (i,j) \in \mathcal{G}.$} between each leg, which is dependent on gait. Under such a change of variables, the coupled canonical systems take the form
\begin{align}
\dot{\x}_i = \f_\x(\x_i,\r) + \sum_{j\in \mathcal{N}_i} \K_{ij} (\R(\phi_{ij})\x_j - \x_i)
\end{align} 
where $\R(\phi_{ij})$ is a rotation matrix of angle $\phi_{ij}$. Three joints per leg with angles $\theta^m_i$ $m\in\{1,2,3\}$ are controlled through transformation systems according to phase-and-reference-based goals $g_\theta^m$ and $g_{\dot{\theta}}^m$
\begin{equation}
\ddot{\theta}_i^{m} = k^m (g_\theta^m(\x_i,\r) - \theta_i^m) + b^m (g_{\dot{\theta}}^m(\x_i, \r) - \dot{\theta}_i^m)
\end{equation}
To approximate these dynamics, torques commanded to the motors are selected as $\tau_i^m = J_i^m \ddot{\theta}_i^m + \tau_{i,fb}^m$ where $J_i^m$ is an estimated motor rotor inertia and $\tau_{i,fb}^m(t)$ allows external feedback coupling from body states. In practice, these feedback torques are formed using a virtual model controller \citep{Pratt01} as in previous work \citep{Ijspeert13}. While their influence is important for the overall control of the balance, the inclusion of these terms is not expressly addressed through the present CPG analysis. Their inclusion in analyzing postural stability represents and important area of future work.

\begin{figure}
\center
\includegraphics[width = .85\columnwidth]{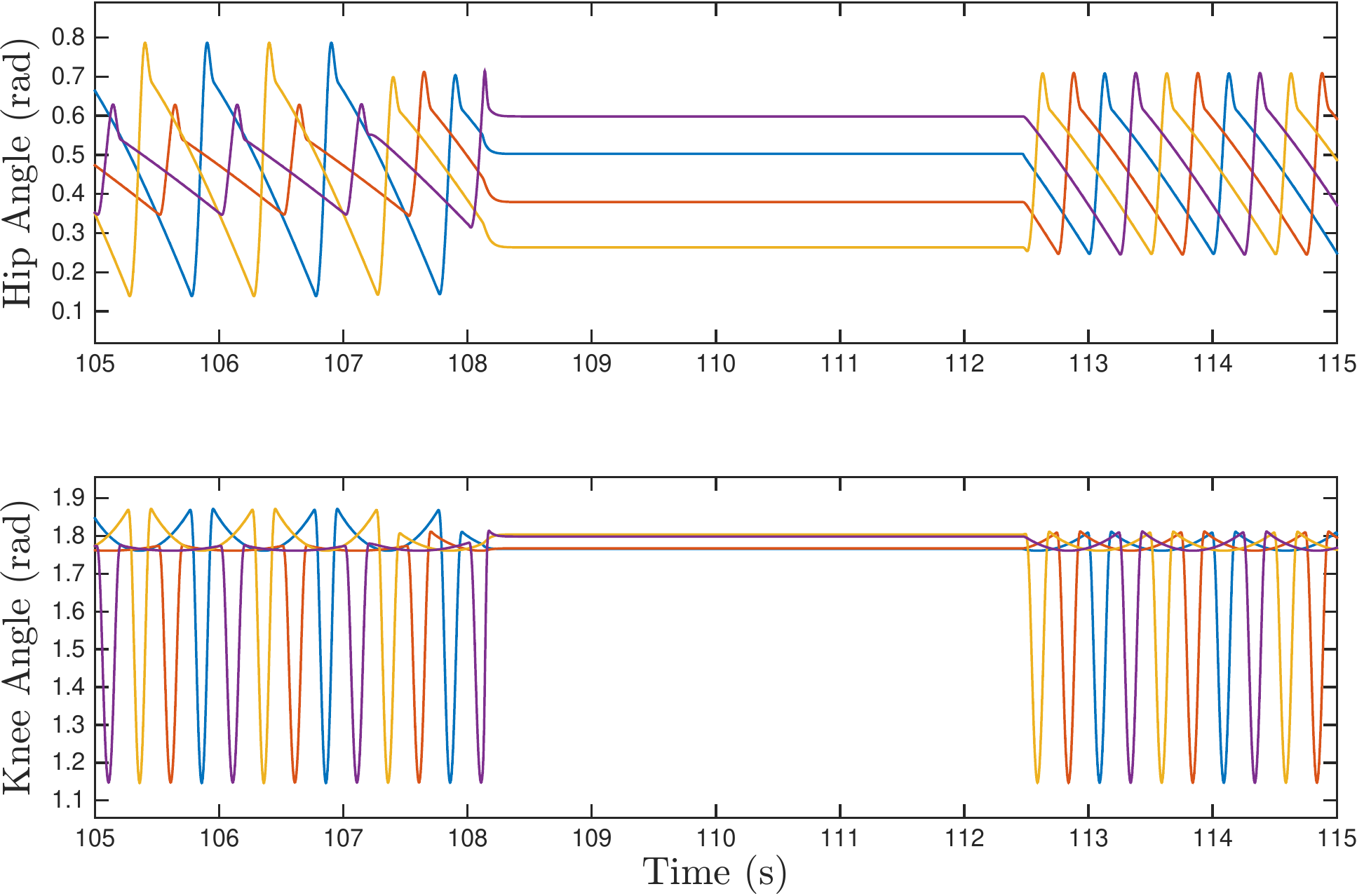}
\caption{Target kinematics for hip and knee joints during an amble gait with Sparsely-Inhibited Rhythmic DMPs.}
\label{fig:TargetKinematics}
\end{figure}

\begin{figure}
\center
\includegraphics[width = .42\columnwidth]{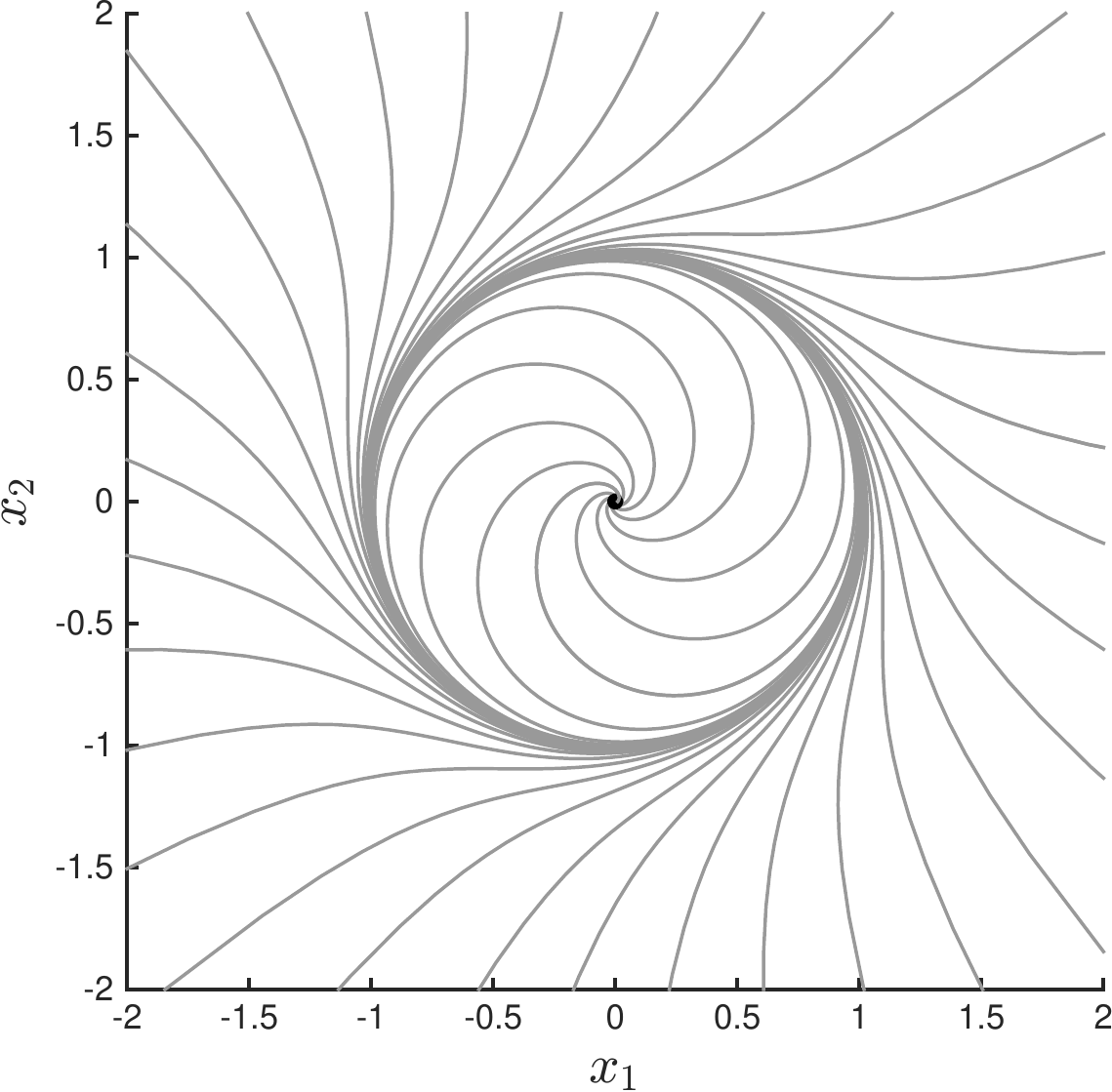}
\includegraphics[width = .42\columnwidth]{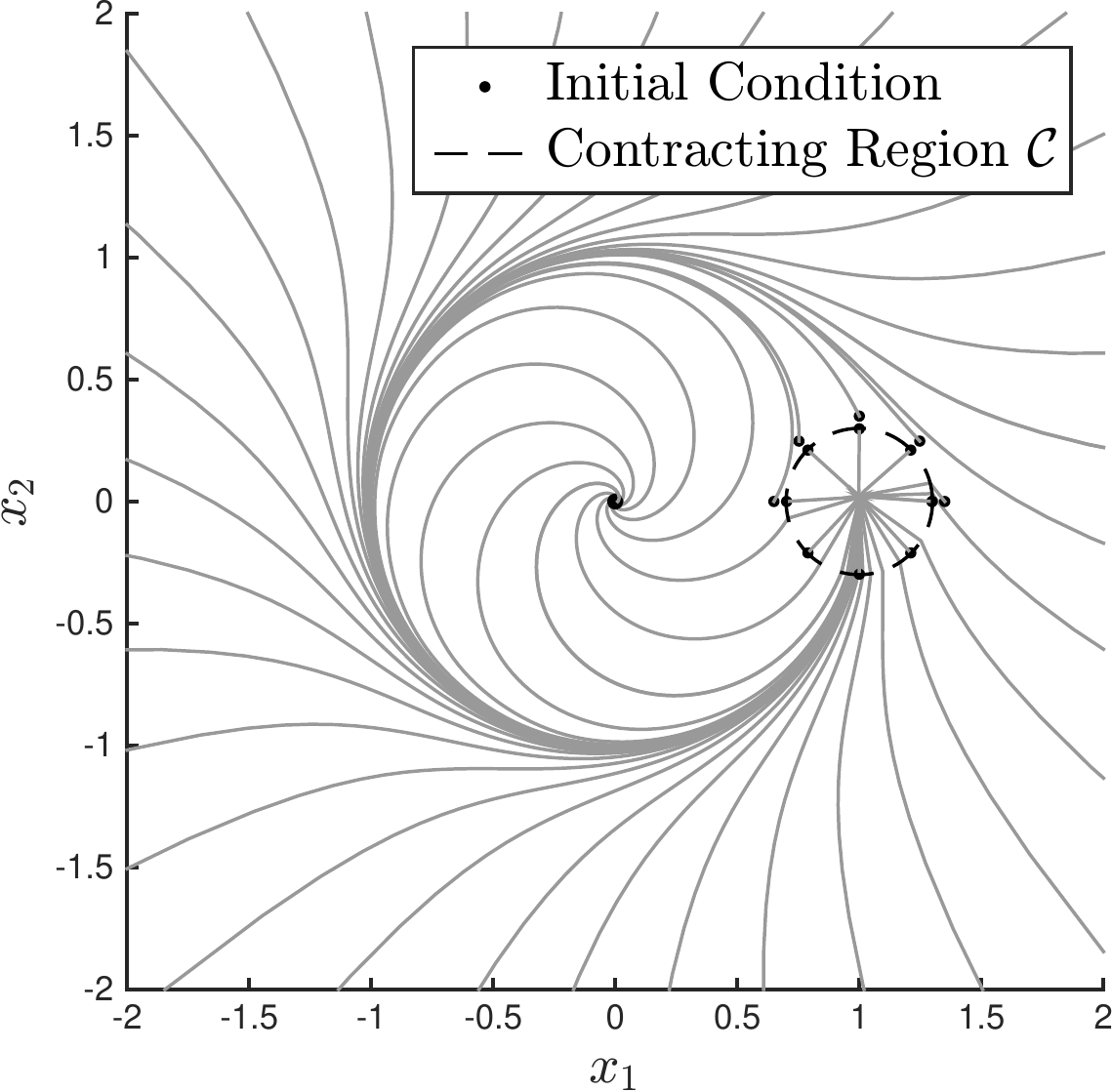}
\caption{Andronov-Hopf oscillator with and without switched inhibition. Upon a state entering a region $\calC$, a strong contracting dynamic is switched, rendering $\calC$ forward invariant and contracting.}
\label{fig:localDynamics}
\end{figure}

\begin{figure}
\center
\includegraphics[width=.85\columnwidth]{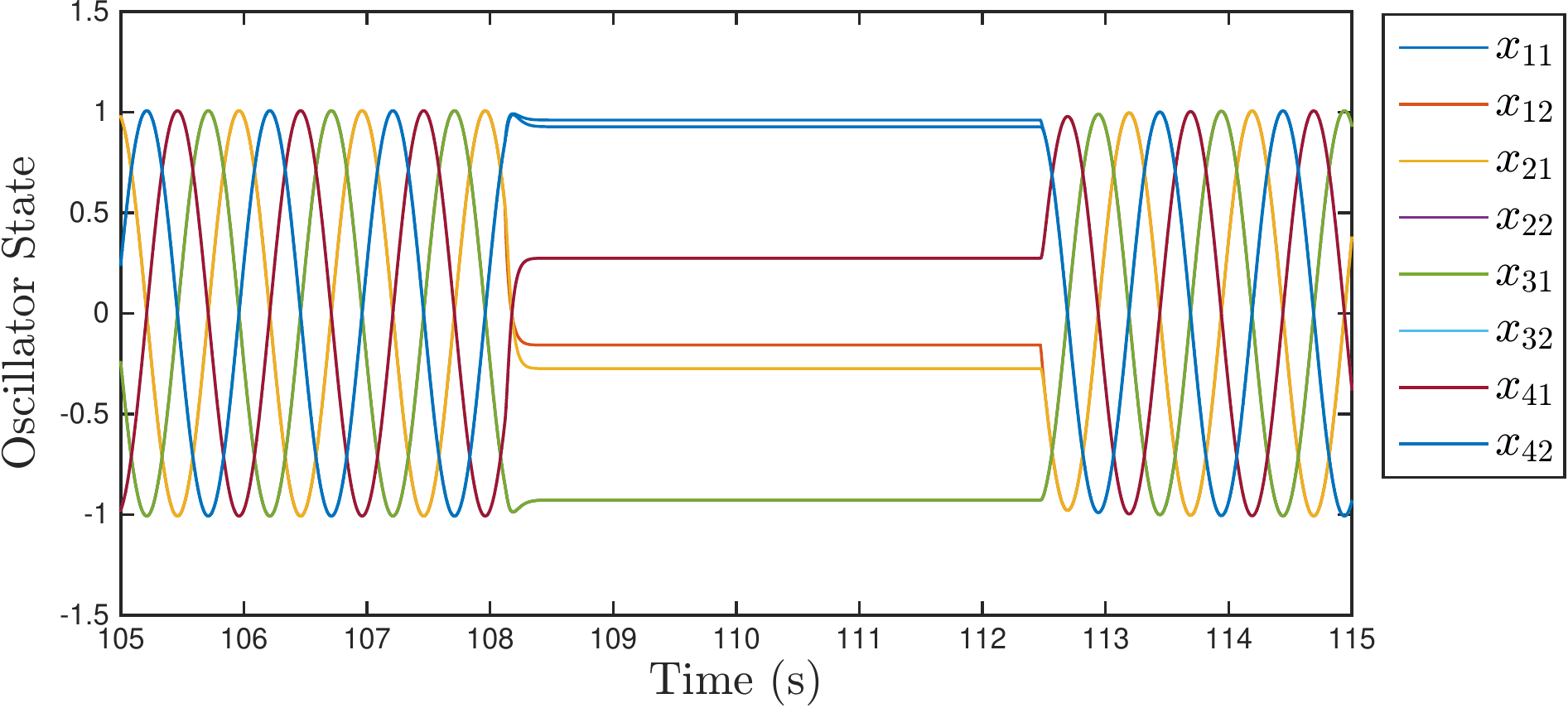}
\caption{Canonical states $\x_i$ across the addition and removal of sparse inhibition to the DMP network.}
\label{fig:OscillatorStates}
\end{figure}

Figure \ref{fig:TargetKinematics} shows the application of Sparsely-Inhibited Rhythmic DMPs to generate an amble gait in the MIT Cheetah. At $t=108.3$s, a contracting dynamic $\g(\x_1) = k_{inh}  ( \x_{1,d} - \x_1 )$ is switched active for states with $| \x_{1,d} - \x_1| \le r_0$ with $\x_{1,d} = [1,\, 0]\T$ and $r_0 = 0.3$. Figure \ref{fig:localDynamics} shows the influence of this contracting dynamic on top of the nominal Andronov-Hopf dynamics. Transverse contraction of the uninhibited network guarantees that the state $\x_1$ reaches the region $\calC$, while switching on $\g(\x_1)$ when $\x_1$ in $\calC$ guarantees contraction of the network. At $t=112.4$s, this sparse influence is disabled, and high-dimensional oscillations resume. Figure \ref{fig:OscillatorStates} shows the canonical states across this transition, with contraction through sparse inhibition, and return to transverse contraction upon removal.

\vspace{-2px}
\section{Discussion}
\vspace{-2px}
A number of additions to these frameworks could be naturally pursued. As legged systems make a break contact with the environment beyond laboratory settings, handling environmental uncertainties is of paramount importance. Contact sensing, while not available in the current experiments, could be integrated to synchronize footfall timings despite touchdown perturbations. The robustness properties of contracting systems \citep{Slotine98} encourage the applicability of this method to handle such inevitable disturbances in unstructured terrains. 

Additional mechanisms to provide coupling from the transformation systems in the canonical dynamics, as in \citep{Chung10}, could be studied to craft responses from disturbances in challenging terrain. Individual limbs could weakly inhibit oscillations, with voting from all limbs required to overcome the threshold of Prop.~\ref{prop:strongEnoughLocalContraction}. More exactly, given dynamics $\g_1,\ldots,\g_N$ contracting in the identity metric, nonnegative inhibition weights $\alpha_1,\ldots,\alpha_N$ could be selected based on feedback such that\begin{equation}
\f_{\rm inh}(\x) = -\L\, \x + [\alpha_1 \g_1(\x_1)^T, \ldots, \alpha_N \g_N(\x_N)^T ]
\nonumber
\end{equation}
in \eqref{eq:inhibitedCanonicalSystem}. Note that the contraction rate of $\f_{\rm inh}$ increases monotonically with each $\alpha_i$, allowing a consensus for inhibition to render $\F_\x(\x,\r) + \f_{\rm inh}$ contracting, even for weak couplings $\L$. Individual gains $\alpha_i$ could be clamped such that an effect of majority voting would be required to inhibit oscillations.  Voting may be pursued within the context of winner take all networks \citep{Douglas10} which admit principled analysis via contraction. More broadly, decentralized feedback could be studied to provide networks of inhibition. Cascades of inhibition are found commonly in the brain \citep{Pfeffer13} perhaps providing a universal building block for complex reasoning akin to NAND-based logic.

	A main challenge in the use of the DMPs for terrain robust legged locomotion rests in addressing the role of body-state feedback in the CPG dynamics. Indeed, the current work does not reason about the contact forces that the limbs are exerting on the world as they move, and it is these contact forces which must be managed to stabilize the body in more challenging scenarios. The modular nature of contraction analysis provides promise that this analysis could be addressed in stages, without immediately requiring high-dimensional verification that is beyond the range of existing tools.

%
%
%
\begin{ack}
\vspace{-10px}
The authors would like to thank Prof. Sangbae Kim for use of the MIT Cheetah robot to conduct experiments.
\end{ack}

\bibliography{ifacconf-extended}            

\begin{thebibliography}{52}
\providecommand{\natexlab}[1]{#1}
\providecommand{\url}[1]{\texttt{#1}}
\providecommand{\urlprefix}{URL }
\expandafter\ifx\csname urlstyle\endcsname\relax
  \providecommand{\doi}[1]{doi:\discretionary{}{}{}#1}\else
  \providecommand{\doi}{doi:\discretionary{}{}{}\begingroup
  \urlstyle{rm}\Url}\fi

\bibitem[{Ajallooeian et~al.(2013{\natexlab{a}})Ajallooeian, Pouya, Sproewitz,
  and Ijspeert}]{Ijspeert13}
Ajallooeian, M., Pouya, S., Sproewitz, A., and Ijspeert, A.J.
  (2013{\natexlab{a}}).
\newblock Central pattern generators augmented with virtual model control for
  quadruped rough terrain locomotion.
\newblock In \emph{IEEE ICRA}, 3321--3328.

\bibitem[{Ajallooeian et~al.(2013{\natexlab{b}})Ajallooeian, van~den Kieboom,
  Mukovskiy, Giese, and Ijspeert}]{Ajallooeian13}
Ajallooeian, M., van~den Kieboom, J., Mukovskiy, A., Giese, M.A., and Ijspeert,
  A.J. (2013{\natexlab{b}}).
\newblock A general family of morphed nonlinear phase oscillators with
  arbitrary limit cycle shape.
\newblock \emph{Physica D}, 263, 41--56.

\bibitem[{Barasuol et~al.(2013)Barasuol, Buchli, Semini, Frigerio, Pieri, and
  Caldwell}]{Semini13}
Barasuol, V., Buchli, J., Semini, C., Frigerio, M., Pieri, E.R.D., and
  Caldwell, D.G. (2013).
\newblock A reactive controller framework for quadrupedal locomotion on
  challenging terrain.
\newblock In \emph{IEEE ICRA}, 2554--2561.

\bibitem[{Bizzi et~al.(1995)Bizzi, Giszter, Loeb, Mussa-Ivaldi, and
  Saltiel}]{Bizzi95}
Bizzi, E., Giszter, S.F., Loeb, E., Mussa-Ivaldi, F.A., and Saltiel, P. (1995).
\newblock Modular organization of motor behavior in the frog's spinal cord.
\newblock \emph{Trends Neurosci}, 18(10), 442--446.

\bibitem[{Burridge et~al.(1999)Burridge, Rizzi, and Koditschek}]{Burridge99}
Burridge, R.R., Rizzi, A.A., and Koditschek, D.E. (1999).
\newblock Sequential composition of dynamically dexterous robot behaviors.
\newblock \emph{Int J Robot Res}, 18(6), 534--555.

\bibitem[{Canter et~al.(2016)Canter, Penney, and Tsai}]{Canter16}
Canter, R.G., Penney, J., and Tsai, L.H. (2016).
\newblock The road to restoring neural circuits for the treatment of
  alzheimer's disease.
\newblock \emph{Nature}, 539(7628), 187--196.
\newblock \urlprefix\url{http://dx.doi.org/10.1038/nature20412}.

\bibitem[{Caughman and Veerman(2006)}]{Caughman06}
Caughman, J.S. and Veerman, J.J.P. (2006).
\newblock Kernels of directed graph laplacians.
\newblock \emph{Elect J Combinatorics}, 13.

\bibitem[{Chen and Slotine(2012)}]{Chen12}
Chen, L. and Slotine, J.J.E. (2012).
\newblock Note on metrics in contraction analysis.
\newblock NSL report, MIT.

\bibitem[{Chung and Slotine(2010)}]{Chung10b}
Chung, S.J. and Slotine, J.J. (2010).
\newblock On synchronization of coupled hopf-kuramoto oscillators with phase
  delays.
\newblock In \emph{IEEE Conf. on Decision and Control}, 3181--3187.

\bibitem[{Chung and Dorothy(2010)}]{Chung10c}
Chung, S.J. and Dorothy, M. (2010).
\newblock Neurobiologically inspired control of engineered flapping flight.
\newblock \emph{Journal of Guidance, Control, and Dynamics}, 33(2), 440--453.

\bibitem[{Dahlquist(1959)}]{Dahlquist59}
Dahlquist, G. (1959).
\newblock Stability and error bounds in the numerical integration of ordinary
  sifferential equations.
\newblock \emph{Trans. Roy. Inst. Tech. Stockholm}, 130.

\bibitem[{Davison et~al.(2016)Davison, Dey, and Leonard}]{Leonard16}
Davison, E.N., Dey, B., and Leonard, N.E. (2016).
\newblock Synchronization bound for networks of nonlinear oscillators.
\newblock In \emph{54th Annual Allerton Conferecence on Communication, COntrol
  and Computing}.

\bibitem[{Desoer and Haneda(1972)}]{Desoer72}
Desoer, C. and Haneda, H. (1972).
\newblock The measure of a matrix as a tool to analyze computer algorithms for
  circuit analysis.
\newblock \emph{IEEE Transactions on Circuit Theory}, 19(5), 480--486.
\newblock \doi{10.1109/TCT.1972.1083507}.

\bibitem[{{G{\'e}rard} and {Slotine}(2006)}]{Gerard06}
{G{\'e}rard}, L. and {Slotine}, J.J. (2006).
\newblock {Neuronal networks and controlled symmetries, a generic framework}.
\newblock \emph{eprint arXiv:q-bio/0612049}.

\bibitem[{Hogan and Sternad(2012)}]{Hogan12}
Hogan, N. and Sternad, D. (2012).
\newblock Dynamic primitives of motor behavior.
\newblock \emph{Biol. Cybernetics}, 106(11-12), 727--739.

\bibitem[{Ijspeert(2008)}]{Ijspeert08}
Ijspeert, A.J. (2008).
\newblock Central pattern generators for locomotion control in animals and
  robots: A review.
\newblock \emph{Neural Networks}, 21(4), 642 -- 653.

\bibitem[{Ijspeert et~al.(2012)Ijspeert, Nakanishi, Hoffmann, Pastor, and
  Schaal}]{Schaal12}
Ijspeert, A.J., Nakanishi, J., Hoffmann, H., Pastor, P., and Schaal, S. (2012).
\newblock Dynamical movement primitives: Learning attractor models for motor
  behaviors.
\newblock \emph{Neural Computation}, 25(2), 328--373.

\bibitem[{Ijspeert et~al.(2002)Ijspeert, Nakanishi, and Schaal}]{Ijspeert02}
Ijspeert, A.J., Nakanishi, J., and Schaal, S. (2002).
\newblock Learning attractor landscapes for learning motor primitives.
\newblock In \emph{Advances in NIPS 15}, 1547--1554. MIT Press.

\bibitem[{Khansari-Zadeh and Billard(2011)}]{Billard11}
Khansari-Zadeh, S. and Billard, A. (2011).
\newblock Learning stable nonlinear dynamical systems with gaussian mixture
  models.
\newblock \emph{IEEE Trans. on Robotics}, 27(5), 943--957.

\bibitem[{Leonov et~al.(1996)Leonov, Burkin, and Shepeljavyi}]{Leonov96}
Leonov, G.A., Burkin, I.M., and Shepeljavyi, A.I. (1996).
\newblock \emph{Frequency Methods in Oscillation Theory}, volume 357 of
  \emph{Mathematics and Its Applications}.
\newblock Springer.

\bibitem[{Lohmiller and Slotine(1998)}]{Slotine98}
Lohmiller, W. and Slotine, J.J.E. (1998).
\newblock On contraction analysis for non-linear systems.
\newblock \emph{Automatica}, 34(6), 683--696.

\bibitem[{Lozinskii(1959)}]{Lozinskii59}
Lozinskii, S.M. (1959).
\newblock Error estimate for numerical integration of ordinary differential
  equations. i,.
\newblock \emph{Izv. Vtssh. Uchebn. Zaved. Mat.}, 5, 222--222.

\bibitem[{Mainen and Sejnowski(1995)}]{Mainen95}
Mainen, Z. and Sejnowski, T. (1995).
\newblock Reliability of spike timing in neocortical neurons.
\newblock \emph{Science}, 268(5216), 1503--1506.

\bibitem[{Manchester et~al.(2015)Manchester, Tang, and Slotine}]{Manchester15}
Manchester, I.R., Tang, J.Z., and Slotine, J.J. (2015).
\newblock Unifying classical and optimization-based methods for robot tracking
  control with control contraction metrics.
\newblock In \emph{Proceedings of ISRR}.

\bibitem[{Manchester and Slotine(2014{\natexlab{a}})}]{Manchester14e}
Manchester, I.R. and Slotine, J.J.E. (2014{\natexlab{a}}).
\newblock {Combination Properties of Weakly Contracting Systems}.
\newblock \emph{ArXiv e-prints}.
\newblock {arXiv}:1408.5174.

\bibitem[{Manchester and Slotine(2014{\natexlab{b}})}]{Manchester14}
Manchester, I.R. and Slotine, J.J.E. (2014{\natexlab{b}}).
\newblock Transverse contraction criteria for existence, stability, and
  robustness of a limit cycle.
\newblock \emph{Sys. \& Control Letters}, 63, 32--38.

\bibitem[{Manchester and Slotine(2015)}]{Manchester15b}
Manchester, I.R. and Slotine, J.E. (2015).
\newblock Control contraction metrics: Convex and intrinsic criteria for
  nonlinear feedback design.
\newblock \emph{CoRR}, abs/1503.03144.
\newblock \urlprefix\url{http://arxiv.org/abs/1503.03144}.

\bibitem[{Marder and Bucher(2001)}]{Marder01}
Marder, E. and Bucher, D. (2001).
\newblock Central pattern generators and the control of rhythmic movements.
\newblock \emph{Current Biology}, 11(23), R986 -- R996.

\bibitem[{Mussa-Ivaldi et~al.(1994)Mussa-Ivaldi, Giszter, and Bizzi}]{Bizzi94}
Mussa-Ivaldi, F.A., Giszter, S.F., and Bizzi, E. (1994).
\newblock Linear combinations of primitives in vertebrate motor control.
\newblock \emph{PNAS}, 91(16), 7534--7538.

\bibitem[{Park et~al.(2017)Park, Wensing, and Kim}]{WensingUnpub16}
Park, H.W., Wensing, P.M., and Kim, S. (2017).
\newblock High-speed bounding with the mit cheetah 2: Control design and
  experiments.
\newblock Submitted to {\em Int J Robot Res}.

\bibitem[{Park et~al.(2015)Park, Wensing, and Kim}]{Park15b}
Park, H.W., Wensing, P., and Kim, S. (2015).
\newblock Online planning for autonomous running jumps over obstacles in
  high-speed quadrupeds.
\newblock In \emph{Proc. of RSS}.

\bibitem[{Pastor et~al.(2009)Pastor, Hoffmann, Asfour, and Schaal}]{Pastor09}
Pastor, P., Hoffmann, H., Asfour, T., and Schaal, S. (2009).
\newblock Learning and generalization of motor skills by learning from
  demonstration.
\newblock In \emph{IEEE ICRA}, 763--768.

\bibitem[{{Perk} and {Slotine}(2006)}]{Slotine06c}
{Perk}, B.E. and {Slotine}, J.J.E. (2006).
\newblock {Motion Primitives for Robotic Flight Control}.
\newblock \emph{eprint arXiv:cs/0609140}.

\bibitem[{Pfeffer et~al.(2013)Pfeffer, Xue, He, Huang, and
  Scanziani}]{Pfeffer13}
Pfeffer, C.K., Xue, M., He, M., Huang, Z.J., and Scanziani, M. (2013).
\newblock Inhibition of inhibition in visual cortex: the logic of connections
  between molecularly distinct interneurons.
\newblock \emph{Nat Neurosci}, 16(8), 1068--1076.

\bibitem[{Pham and Slotine(2007)}]{Pham07}
Pham, Q.C. and Slotine, J.J. (2007).
\newblock Stable concurrent synchronization in dynamic system networks.
\newblock \emph{Neural Networks}, 20(1), 62 -- 77.

\bibitem[{Pratt et~al.(2001)Pratt, Chew, Torres, Dilworth, and Pratt}]{Pratt01}
Pratt, J., Chew, C.M., Torres, A., Dilworth, P., and Pratt, G. (2001).
\newblock Virtual model control: An intuitive approach for bipedal locomotion.
\newblock \emph{Int J Robot Res}, 20(2), 129--143.

\bibitem[{Ravichandar and Dani(2015)}]{Ravichandar15}
Ravichandar, H. and Dani, A. (2015).
\newblock Learning contracting nonlinear dynamics from human demonstration for
  robot motion planning.
\newblock In \emph{Proceedings of DSCC}. ASME.

\bibitem[{Rohrer et~al.(2004)Rohrer, Fasoli, Krebs, Volpe, Frontera, Stein, and
  Hogan}]{Hogan04}
Rohrer, B., Fasoli, S., Krebs, H.I., Volpe, B., Frontera, W.R., Stein, J., and
  Hogan, N. (2004).
\newblock Submovements grow larger, fewer, and more blended during stroke
  recovery.
\newblock \emph{Motor Control}, 8(4), 472--483.

\bibitem[{Russo et~al.(2013)Russo, di~Bernardo, and Sontag}]{Sontag13}
Russo, G., di~Bernardo, M., and Sontag, E.D. (2013).
\newblock A contraction approach to the hierarchical analysis and design of
  networked systems.
\newblock \emph{IEEE Transactions on Automatic Control}, 58(5), 1328--1331.
\newblock \doi{10.1109/TAC.2012.2223355}.

\bibitem[{Russo and Slotine(2011)}]{Russo11}
Russo, G. and Slotine, J.J.E. (2011).
\newblock Symmetries, stability, and control in nonlinear systems and networks.
\newblock \emph{Phys. Rev. E}, 84, 041929.
\newblock \doi{10.1103/PhysRevE.84.041929}.

\bibitem[{Rutishauser et~al.(2010)Rutishauser, Douglas, and
  Slotine}]{Douglas10}
Rutishauser, U., Douglas, R.J., and Slotine, J.J. (2010).
\newblock Collective stability of networks of winner-take-all circuits.
\newblock \emph{Neural Computation}, 23(3), 735--773.

\bibitem[{Schaal(2006)}]{Schaal06}
Schaal, S. (2006).
\newblock Dynamic movement primitives-a framework for motor control in humans
  and humanoid robotics.
\newblock In \emph{Adaptive Motion of Animals and Machines}, 261--280.
  Springer.

\bibitem[{Seo et~al.(2010)Seo, Chung, and Slotine}]{Chung10}
Seo, K., Chung, S.J., and Slotine, J.J. (2010).
\newblock Cpg-based control of a turtle-like underwater vehicle.
\newblock \emph{Autonomous Robots}, 28(3), 247--269.

\bibitem[{Simon(1962)}]{Simon62}
Simon, H.A. (1962).
\newblock The architecture of complexity.
\newblock \emph{Proc. of the American Philosophical Society}, 106(6), 467--482.

\bibitem[{Singh et~al.(2017)Singh, Majumdar, Slotine, and Pavone}]{Singh17}
Singh, S., Majumdar, A., Slotine, J.J., and Pavone, M. (2017).
\newblock Robust online motion planning via contraction theory and convex
  optimization.
\newblock In \emph{ICRA submission}.

\bibitem[{Slotine and Lohmiller(2001)}]{Slotine01}
Slotine, J.J. and Lohmiller, W. (2001).
\newblock Modularity, evolution, and the binding problem: a view from stability
  theory.
\newblock \emph{Neural Networks}, 14(2), 137 -- 145.

\bibitem[{Tang and Manchester(2014)}]{Manchester14b}
Tang, J. and Manchester, I. (2014).
\newblock Transverse contraction criteria for stability of nonlinear hybrid
  limit cycles.
\newblock In \emph{IEEE Conf. on Decision and Control (CDC)}, 31--36.

\bibitem[{Tedrake et~al.(2010)Tedrake, Manchester, Tobenkin, and
  Roberts}]{Tedrake10}
Tedrake, R., Manchester, I.R., Tobenkin, M., and Roberts, J.W. (2010).
\newblock {LQR}-trees: Feedback motion planning via sums-of-squares
  verification.
\newblock \emph{Int J Robot Res}, 29(8), 1038--1052.

\bibitem[{Vidyasagar(2002)}]{Vidyasagar}
Vidyasagar, M. (2002).
\newblock \emph{Nonlinear Systems Analysis}.
\newblock Classics in Applied Mathematics. SIAM.

\bibitem[{Wang and Slotine(2005)}]{Slotine05}
Wang, W. and Slotine, J.J.E. (2005).
\newblock On partial contraction analysis for coupled nonlinear oscillators.
\newblock \emph{Biological Cybernetics}, 92(1), 38--53.

\bibitem[{Wensing et~al.(2016)Wensing, Wang, Seok, Otten, Lang, and
  Kim}]{WensingUnpub15}
Wensing, P.M., Wang, A., Seok, S., Otten, D., Lang, J., and Kim, S. (2016).
\newblock Proprioceptive actuator design in the {MIT} cheetah: Impact
  mitigation and high-bandwidth physical interaction for dynamic legged robots.
\newblock Submitted to {\em IEEE Trans. on Robotics}.

\bibitem[{Williamson(1999)}]{Williamson99}
Williamson, M.M. (1999).
\newblock \emph{Robot Arm Control Exploiting Natural Dynamics}.
\newblock Ph.D. thesis, MIT.

\end{thebibliography}

\appendix

\section{Selected Proofs}    

\subsection{Proof of Theorem 1}

A partial sketch of this result was originally offered in \citep{Manchester14}.

By the results of \citet{Chen12}, if a system is transverse contracting with rate $\lambda$, there exists a singular metric $\M_s$ with rank $n-1$ such that $\M_s \f = 0$ and
\begin{equation}
\dot{\M}_s + \A\T \M_s + \M_s \A \le -2 \lambda \M_s\,.
\end{equation}
Such a solution is given by
\[
\M_s = \int_0^\infty \V(t,\x)\T \Q(\x(t)) \V(t,\x) {\rm d}t
\]
where $\Q(\x)=\Q(\x)\T\ge0$,  $\Q(\x)$ bounded, ${\rm rank}(\Q(\x))=n-1$, and $\Q(\x) \f(\x)=0$ over the transverse contraction region. $\V(t,\x)$ is the fundamental matrix of the linear time varying system
\[
\dot{\v} = \underbrace{\left( \A - \frac{\f \f\T}{\f\T\f} (\A + \A\T) \right)}_{\A_\v} \v\,.
\]
That is $\V(0,\x) = \I$, and $\dot{\V}(t,\x) = \A_{\v}(t,\x) \V(t,\x)$. The singular metric $\M_s$ is used as a starting point towards a full-rank metric with the desired eigenstructure on an associated generalized Jacobian.

Letting $\pib(\x) = \f\T/\abs{\f}$ and $\Pib(x) \in \mathbb{R}^{ (n-1) \times n}$ complete a smooth orthonormal basis, it follows that $\M_s$ can be written as
\[
\M_s = \Pib\T(\x)\, \tilde{\M}_s(\x) \, \Pib(\x) 
\]
for some symmetric positive definite $\tilde{\M}_s(\x) \in \mathbb{R}^{ (n-1) \times (n-1)}$.
 
We note that the differential dynamics satisfy
\[
\dot{\delt{x}} = \A(t,\x) \delt{x}
\]
and further observe that $\delt{x}(t) = \f(\x(t))$ is a solution to the differential dynamics. Defining the differential change of variables
\[
\begin{bmatrix} \delt[1]{z} \\ \delt[2]{z} \end{bmatrix} = \underbrace{\begin{bmatrix} \frac{\f\T}{|\f|^2} \\[2ex] \tilde{\M}_s^{1/2} \Pib \end{bmatrix}}_{\Thetab_{\x}} \delt{x}
\]
Since $\delt{x}(t) = \f(\x(t))$ a solution to the differential dynamics, it follows that $\delt[1]{z}(t) \equiv 1$, $\delt[2]{z}(t) \equiv 0$ is a solution to the different differential dynamics for $\delt{z}$. Thus
\[
\ddt \begin{bmatrix} \delt[1]{z} \\ \delt[2]{z} \end{bmatrix} = \begin{bmatrix} 0 & \A_{12}(t,\x) \\ 0 & \A_{22}(t,\x) \end{bmatrix} \begin{bmatrix} \delt[1]{z} \\ \delt[2]{z} \end{bmatrix}
\]
We further observe that, by construction,
\begin{align}
\ddt \delt[2]{z}\T\, \delt[2]{z} &= \ddt \delt{x}\T \M_s \delt{x} \\
	  						&= \delt{x}\T \left(\dot{\M}_s + \A\T \M_s + \M_s \A \right) \delta x \\
						    &\le -2 \lambda \delt{x}\T \M_s \delt{x} \\
						    &= -2 \lambda \delt[2]{z}\T\, \delt[2]{z}\,. \label{eq:convergentz2}
\end{align}
Thus, the subspace of differentials from $\delt[2]{z}$ possess a contracting dynamic which drives the indifferent $\delt[1]{z}$ subsystem. While this generalized Jacobian for $\delt{z}$ has eigenvalues with the desired structure, the symmetric part of this generalized Jacobian does not. To obtain the desired eigenstructure on the symmetric part of the generalized Jacobian, a further state transformation is pursued through construction of a new metric for $\delta \z$.

We consider the following structure for a metric over $\delta \z$:
\[
\M_z(\x) = \begin{bmatrix} 1  & \M_{21}\T(\x) \\ \M_{21}(\x) & \M_{22}(\x) \end{bmatrix} \,.
\] 
The rate of change in differential length is given as
\begin{align*}
\ddt \left( \delt{z}\T \M_z \delt{z} \right)&= \ddt \left(  \delt[1]{z}\T\,\delt[1]{z} \right) +
					      \ddt \left(  \delt[2]{z}\T\,\M_{22} \delt[2]{z} \right) \\
					      & \phantom{=} +\ddt 2 \left( \delt[2]{z}\T \M_{21} \delt[1]{z} \right)\\
					   &= \delt[2]{z}\T \left(\dot{\M}_{22} + \M_{22} \A_{22} + \A_{22}\T \M_{22} \right. \\ & \phantom{=} \quad\quad\quad~~\left.+  \M_{21} \A_{12} + \A_{12}\T \M_{21}\T \right) \,\delt[2]{z}  \\
					   & \quad
					    + 2 \delt[2]{z}\T \left(\dot{\M}_{21} + \A_{22}\T \M_{21} + \A_{12}\T \right) \delt[1]{z}
\end{align*}

We'll first attempt to determine a solution $\M_{21}(\x)$  and then for $\M_{22}(\x)$. Towards canceling cross terms above, we seek a solution to 
\begin{equation}
\dot{\M}_{21} + \A_{22}\T \M_{21} + \A_{12}\T = 0
\label{eq:desContrIdentity}\,.
\end{equation}

\begin{proposition}
A solution to (\ref{eq:desContrIdentity}) is given by 
\[
\M_{21}(\x) = \int_0^\infty \U_2\T(\tau,\x) \A_{12}\T(\tau,\x) \d \tau
\]
with
\[
\ddt \U_2(t,\x) = \A_{22}(t,\x)\, \U_2(t,\x)\quad \U_2(0,\x) = \I  
\]
\end{proposition}
\begin{pf}
By construction. Equation \ref{eq:desContrIdentity} is equivalent to
\begin{align}
&\U_2\T(t,\x) \left( \A_{22}\T(t,\x) \M_{21}(t,\x) + \dot{\M}_{21}(t,\x) \right) \\
=& -\U_2\T(t,\x) \A_{12}\T(t,\x)
\end{align}
which in turn is equivalent to
\[
\ddt\left( \U_2\T(t,\x) \, \M_{21}(t,\x) \right) = -\U_2
\T(t,\x) \A_{12}\T(t,\x)\,.
\]
Integrating both sides over the interval $(0,\infty)$ provides:
\[
-\U_2\T(0,\x) \,\M_{21}(0,\x) = \int_0^\infty -\U_2\T(t,\x) \,\A_{12}\T(t,\x) \d t
\]
The right hand side converges since, due to (\ref{eq:convergentz2}), $\|\U_2(t,\x)\| \le C {\rm{e}}^{-\lambda t}$ for some $C>0$.\hfill$\qed$
\end{pf}

Letting $\M_{21}$ as prescribed:
\begin{align}
\ddt \left( \delt{z}\T \M_z \delt{z} \right)&= \delt[2]{z}\T \left(\dot{\M}_{22} + \M_{22} \A_{22} + \A_{22}\T \M_{22}+\right.  \nonumber\\ &\phantom{=}\left.~~~~~~~~~   \M_{21} \A_{12} + \A_{12}\T \M_{21}\T \right) \,\delt[2]{z}
\label{eq:ForDesiredRank}
\end{align}
Letting $\Q=\Q\T>0$ and $r < \lambda$ we form $\M_{22}$ by solving the differential equation:
\begin{align}
\dot{\M}_{22}+ \M_{22} \A_{22} + \A_{22}\T \M_{22}+ 2 r \M_{22}~& \nonumber\\ +  \M_{21} \A_{12} + \A_{12}\T \M_{21}\T + \Q &= 0
\label{eq:desContrIdentity2}
\end{align}
\begin{proposition}
A solution to Equation \ref{eq:desContrIdentity2} is given by
\[
\M_{22}(\x) = \int_0^\infty {\rm e}^{2 r t} \U_2\T \left( \M_{21} \A_{12} + \A_{12}\T \M_{21}\T + \Q \right) \U_2\, \d t
\]
with each shorthand $\M_{21}=\M_{21}(t,\x)$, $\A_{12}:=\A_{12}(t,\x)$, $\U_2:=\U_2(t,\x)$. 
\end{proposition}

\begin{pf}
Analogous to the solution for Equation \ref{eq:desContrIdentity},  Equation \ref{eq:desContrIdentity2} is identical to requiring
\begin{align}
&\ddt \left( {\rm e}^{2rt} \U_2(t,\x)\T \M_{22}(t,\x) \U_2(t,\x) \right ) 
\\&~= -{\rm e}^{2 r t} \U_2(t,\x)\T \left(\M_{21} \A_{12} + \A_{12}\T \M_{21}\T + \Q \right) \U_2(t,\x)
\end{align}
Integrating both sides over the interval $(0,\infty)$ again provides the desired result.\hfill$\qed$
\end{pf}

\begin{remark}
In order to ensure $\M_z > 0$ it follows that $\M_{22}$ must satisfy  $\M_{22} > \M_{21} \M_{21}\T$. $\Q$ can be scaled by a suitable factor to meet this requirement without loss of generality to the previous development.
\end{remark}

Putting these ingredients together, it follows that
\begin{align}
\ddt \left( \delt{z}\T \M_z \delt{z} \right) = -\delt[2]{z}\T\left( \Q + 2 r \M_{22}\right) \,\delt[2]{z} 
\label{eq:finalineq}
\end{align}
A smooth factorization of $\M_z= \Thetab_z^T \Thetab_z$ finally gives rise to a subsequent change of differential coordinates $\delt{y} := \Thetab_z \delt{z}$. Letting $\Thetab:=\Thetab_z \Thetab_x$, from  (\ref{eq:finalineq}) it follows
\begin{align}
\ddt \left( \delt{x}\T \Thetab\T \Thetab \delt{x} \right)
=& \ddt \left( \delt{y}\T \delt{y} \right)  \\
=& 2 \delt{y}\T \F_s \delt{y} \\
=& -\delt[2]{z}\T\left( \Q + 2 r \M_{22}\right) \,\delt[2]{z}
\end{align}
Thus, $\F_s$ is negative semidefinite and has rank $n-1$.

\subsection{Disturbed Transverse Contracting Systems}
\label{sec:disturbed}

\begin{proposition}
Suppose \eqref{eq:sys} autonomous, transverse contracting on a compact $\calK$ with rate $\lambda$ under metric $\M(\x)$. Let $\x(t)$ the solution from some initial condition $\x_0$. 
Suppose a solution $\x_d(t)$ to the disturbed system
\[
\dot \x_d = \f(\x_d) + \w(t)\,
\]
from the same initial condition $\x_0$. Suppose $\w$ uniformly bounded $| \w(t)| \le \overline{w}$ and $\x_d(t) \in \cal{K}$ for all possible realizations of $\w(\cdot)$. Then $\forall t>0$, $\inf_\tau| \x_d(t) - \x(\tau)| \le \tfrac{R}{ \lambda} \overline{w}$, where $R>0$ depends only on $\M$. 
\end{proposition}

To prove the result, we will argue the existence of a control $u(t)$ to the virtual system 
\begin{equation}
\label{eq:ccmfortc}
\dot \y = \f(\y) u
\end{equation}
with initial condition $\y(0) = \x_0$ such that  $| \x_d(t) - \y(t)| \le \tfrac{R}{ \lambda} \overline{w}$. Note that conditions on $\M$ being a control contraction metric for \eqref{eq:ccmfortc} are necessary and sufficient for $\M$ to be a transverse contraction metric for \eqref{eq:sys} \citep{Manchester15b}.

Towards a proof of this result, let 
\begin{align}
&\gammab(\x_1, \x_2):\{ \gammab(\cdot) \in \calC^{\infty}([0,1], \calK) \textrm{~s.t.~} \frac{\partial}{\partial s} \gammab(s)\ne0 \nonumber\\&\quad\quad\quad\quad~~~~~~~~~~\forall s\in (0,1) \textrm{, }  \gammab(0) = \x_1,~ \gammab(1)=\x_2\} \nonumber
\end{align}
the set of smooth paths in $\calK$ from $\x_1$ to $\x_2$. Further, let
\begin{equation}
d(\x_1,\x_2) := \inf_{\gamma \in \gammab(\x_1,\x_2)} \int_0^1 \sqrt{ \gammab_s(s)^T \M(\gammab(s)) \gammab_s(s)} \textrm{d}s
\nonumber
\end{equation}
the Riemann distance, where $\gammab_s := \pd{s} \gammab$. 
Similarly, let
\begin{equation}
e(\x_1,\x_2) := \inf_{\gamma \in \gammab(\x_1,\x_2)} \int_0^1  \gammab_s(s)^T \M(\gammab(s)) \gammab_s(s) \textrm{d}s \nonumber
\end{equation}
the Riemann energy satisfying $e(\x_1,\x_2)= d(\x_1,\x_2)^2$.

\begin{pf}
Let $\gammab(\cdot)$ the geodesic between $\x_d(t)$ and $\y(t)$ at some time $t$. From the formula for the first variation of energy \citep{Manchester15b}
\begin{align}
\frac{1}{2} D^+ e( \x_d(t), \y(t) ) =& \gammab_s(0)^T \M(\x_d) (\f(\x_d) + \w) \nonumber \\
									 &- \gammab_s(1)^T \M(\y) \f(\y) u \label{eq:lagrange}
\end{align}
where $D^+$ denotes the upper Dini derivative.  The control contraction metric allows the Riemannian energy to be effectively used as a control Lyapunov function. In this light, the control contraction metric conditions imply a Artstein/Sontag CLF condition \citep{Manchester15b} that if $\gammab_s(1)^T \M(\y) \f(\y) = 0$ then $\gammab_s(0)^T \M(\x_d) \f(\x_d) < -\lambda e(\x_d, \y)$. 

It follows that at each time, there exists $u$ such that 
\[
\gammab_s(0)^T \M(\x_d) \f(\x_d) - \gammab_s(1)^T \M(\y) \f(\y) u < -\lambda e(\x_d, \y)
\]
Suppose $\M = \Thetab^T \Thetab$ and let $\boldsymbol{\delta}(s)  = \Thetab(\gammab(s)) \gammab_s(s)$. Since the velocity field of a geodesic is parallel along the geodesic, $e(\x_d, \y) = \gammab_s(s)^T \M(\gammab(s)) \gammab_s(s)$, $\forall s \in [0,1]$. This further implies $| \boldsymbol{\delta}(s)| = d(\x_d, \y)$ \citep{Singh17}. Under this control, and through application of the Cauchy-Schwarz inequality \eqref{eq:lagrange} provides
\begin{align}
\frac{1}{2} D^+ e( \x_d(t), \y(t) ) \le & d(\x_d(t),\y(t)) |\Thetab( \x_d(t)) \w(t)|_2 \nonumber\\
									   &- \lambda e(\x_d(t),\y(t))
\end{align}
Letting $\overline{\theta} = \sup_{\x \in \calK} \| \Thetab(\x) \|$ and $\overline{w}=\sup_t | w(t)|$, it follows from the comparison lemma that $d(\x_d(t), \y(t)) \le \tfrac{\overline{\theta}}{\lambda} \overline{w}$ $\forall t \ge 0$. Suppose $\underline{\theta}>0$ such that $\underline{\theta}^2 \I \le \Thetab^T \Thetab$. Then $\underline{\theta} |\x_d(t) - \y(t)| \le d(\x_d(t), \y(t))$. It finally follows that with $R = {\overline{\theta}}/{\underline{\theta}}$, $|\x_d(t) - \y(t)| \le \tfrac{R}{\lambda} \overline{w}$. Note that $R$ is an upper bound on the condition number of $\Thetab$. \hfill $\qed$

\end{pf}

\end{document}